\documentclass[conference]{IEEEtran}
\IEEEoverridecommandlockouts
\usepackage[noadjust]{cite}
\usepackage{amsmath,amssymb,amsfonts}
\usepackage{algorithmic}
\usepackage{graphicx}
\usepackage{textcomp}
\usepackage{xcolor}
\usepackage{multirow}
\usepackage[letterpaper,right=0.63in,left=0.63in,top=0.75in,bottom=1.11in]{geometry}

\usepackage{algorithm} 
\usepackage{algorithmic} 

\newtheorem{theorem}{Theorem}
\newtheorem{definition}{Definition}
\newtheorem{lemma}{Lemma}
\newtheorem{corollary}{Corollary}
\newtheorem{example}{Example}

\newtheorem{remark}{Remark}
\newtheorem{observation}{Observation}

\def\BibTeX{{\rm B\kern-.05em{\sc i\kern-.025em b}\kern-.08em
    T\kern-.1667em\lower.7ex\hbox{E}\kern-.125emX}}
\begin{document}

\title{The Impact of the Distance Between Cycles on Elementary Trapping Sets\\
\thanks{This work was supported by the National Key R\&D Program of China (No. 2023YFA1009602).}
}

\author{
  \IEEEauthorblockN{Haoran Xiong\IEEEauthorrefmark{2}\IEEEauthorrefmark{3}, 
   Guanghui Wang\IEEEauthorrefmark{4},
   Zhiming Ma\IEEEauthorrefmark{2}\IEEEauthorrefmark{3}, 
   and Guiying Yan\IEEEauthorrefmark{2}\IEEEauthorrefmark{3}\IEEEauthorrefmark{1}}

   \IEEEauthorblockA{\IEEEauthorrefmark{2}
                     University of Chinese Academy of Sciences}

   \IEEEauthorblockA{\IEEEauthorrefmark{3}
                     Academy of Mathematics and Systems Science, CAS}

   \IEEEauthorblockA{\IEEEauthorrefmark{4}
                     School of Mathematics, Shandong University}

    Email: 
	xionghaoran@amss.ac.cn,
	ghwang@sdu.edu.cn, mazm@amt.ac.cn, yangy@amss.ac.cn
    }

\maketitle

\begin{abstract}
    Elementary trapping sets (ETSs) are the main culprits of the performance of low-density parity-check (LDPC) codes in the error floor region. Due to their large quantities and complex structures, ETSs are difficult to analyze. This paper studies the impact of the distance between cycles on ETSs, focusing on two special graph classes: theta graphs and dumbbell graphs, which correspond to cycles with negative and non-negative distances, respectively. We determine the Tur\'an numbers of these graphs and prove that increasing the distance between cycles can eliminate more ETSs. Additionally, using the linear state-space model and spectral theory, we prove that increasing the length of cycles or distance between cycles decreases the spectral radius of the system matrix, thereby reducing the harmfulness of ETSs. This is consistent with the conclusion obtained using Tur\'an numbers. For specific cases when removing two 6-cycles with distance of -1, 0 and 1, respectively, we calculate the sizes, spectral radii, and error probabilities of ETSs. These results confirm that the performance of LDPC codes improves as the distance between cycles increases. Furthermore, we design the PEG-CYCLE algorithm, which greedily maximizes the distance between cycles in the Tanner graph. Numerical results show that the QC-LDPC codes constructed by our method achieve performance comparable to or even superior to state-of-the-art construction methods.
\end{abstract}


\section{Introduction}\label{sec:introduction}

As an important class of modern coding theory, low-density parity-check (LDPC) codes have received widespread attention for their excellent error-correction capabilities and efficient parallel iterative decoding algorithm~\cite{gallager1962low,tanner1981recursive,mackay1997near}. As capacity-approaching codes, LDPC codes are extensively used in various systems, such as Wi-Fi, optical communication, microwave systems, and data storage~\cite{chung2001design}. Among them, quasi-cyclic LDPC (QC-LDPC) codes are particularly important due to their efficient hardware implementation, making them a popular choice in standards such as 5G NR~\cite{5gnr, townsend1967self, okamura2003designing}. 

The error rate curve of LDPC codes under iterative decoding is typically divided into two regions based on channel quality, which is usually characterized by the signal-to-noise ratio (SNR): the waterfall region and the error floor region. The error floor region occurs at high SNR values and is characterized by a gradual reduction in error rate as the channel quality improves, with a noticeable change in the slope of the error rate curve.

A Tanner graph, which is a bipartite graph composed of variable nodes and check nodes, corresponds one-to-one to an LDPC code and plays an important role in the decoding process~\cite{tanner1981recursive,davey1998low}. 
A Tanner graph is called variable-regular if all its variable nodes have the same degree, denoted by $d_L(v)$.

The error floor behavior of an LDPC code is primarily caused by specific graph structures in the Tanner graph, known as trapping sets~\cite{richardson2003error}. An $(a,b)$ trapping set (TS) is defined as an induced subgraph containing $a$ variable nodes and $b$ check nodes with odd degrees, along with any number of check nodes with even degrees in the Tanner graph. Here, $a$ is called the size of the TS. An elementary trapping set (ETS) is a specific type of TS where all check nodes have degrees of either 1 or 2 and is considered the most harmful among TSs~\cite{cole2006general, milenkovic2006asymptotic}. 

Removing certain ETSs in the Tanner graph is important for improving the performance of LDPC codes in the error floor region,
However, identifying the non-isomorphic structures of ETSs with varying values of $a$ and $b$ is challenging. Moreover, McGregor \emph{et al.}~\cite{mcgregor2010hardness} proved that determining the minimum size of ETSs in a Tanner graph is NP-hard. 
Given these difficulties in directly analyzing ETSs, researchers primarily focused on structural properties that directly influence ETSs. The most widely studied approaches include optimizing the minimum distance $d_{min}$~\cite{kou2001low, huang2011trapping,mittelholzer2002efficient,battaglioni2020improving} and increasing the girth $g$ of the Tanner graph~\cite{mitchell2014quasi, tasdighi2016efficient, tadayon2018efficient, mo2020designing, wang2008construction, wang2013hierarchical,bocharova2011searching}. Additionally, other methods have been proposed for removing ETSs, such as removing specific graph structures and particular TSs~\cite{kelley2008ldpc,tao2017construction,naseri2019construction,amirzade2021qc,hashemi2015characterization, karimi2014characterization,amirzade2022protograph,xiong2024theoretical}. 

In this paper, we introduce a new structural property: the distance between cycles, for analyzing and eliminating ETSs. This idea not only complements existing techniques but also offers a more effective way to improve the performance of LDPC codes in the error floor region.

\subsection{Previous Work}

In~\cite{hashemi2015characterization, karimi2014characterization}, Banihashemi \emph{et al.} used computer programming to substantiate that ETSs with relatively small values of $a$ and $b$ can be generated by short cycles or non-cycle graphs, which characterize the basic structures of ETSs. 
By avoiding the 8-cycle with a chord in the Tanner graph, Amirzade \emph{et al.}~\cite{amirzade2022protograph} constructed QC-LDPC codes with girth $g=6$. These codes are free of all $(a,b)$-ETSs where $a\leq5$ and $b\leq3$ for $d_L(v)=3$, and $a\leq7$ and $b\leq4$ for $d_L(v)=4$. Furthermore, they derived an inequality $b\geq a\gamma-\frac{2a^3}{4a-3}$ for $(a,b)$-ETSs in the Tanner graphs with a variable-regular degree $d_L(v)=\gamma$. 
In~\cite{xiong2024theoretical}, the authors improved this bound and considered the case of girth 8. By restricting the number of common edges between two 8-cycles, they eliminated $(a,b)$-ETSs with $a\leq 7$, $b<3$ when $d_L(v)=3$ and $(a,b)$-ETSs with $a\leq 9$, $b<6$ when $d_L(v)=4$.
We observe that removing the 8-cycle with a chord in the Tanner graph~\cite{amirzade2022protograph} and restricting the number of common edges between two 8-cycles~\cite{xiong2024theoretical} imply the elimination of cycles with small distances.

\subsection{Our Contributions}

In this paper, we focus on the ETSs in variable-regular Tanner graphs with $d_L(v)=\gamma$. We use the variable node (VN) graph~\cite{koetter2004pseudo} to represent an ETS, where a one-to-one correspondence exists between them in the case of variable-regular Tanner graphs. Here, we consider two special classes of graphs—theta graphs and dumbbell graphs—which form the fundamental structures of VN graphs for ETSs. We establish the relationship between the removal of specific substructures in the VN graph and constraints on ETSs in the Tanner graph through Tur\'an numbers. As a cornerstone of graph theory, Tur\'an numbers provide profound insights into how local constraints influence global graph structures. From the perspective of Tur\'an numbers, we prove that removing structures with smaller Tur\'an numbers in the VN graph can eliminate more ETSs in the Tanner graph.

To further analyze the behavior of these structures in ETSs, we use the linear state-space model~\cite{Butler2014errorfloor}. By spectral theory, we prove that increasing the cycle lengths or their distances reduces the spectral radius of the corresponding system matrix. This conclusion is supported by numerical examples. By combining this finding with the spectral radius analysis in~\cite{Butler2014errorfloor}, we conclude that increasing the cycle lengths and the distances improves the performance of LDPC codes, which is consistent with the conclusion obtained from Tur\'an numbers. 
Furthermore, we utilize the linear state-space model to simulate the rate of error probability reduction of ETSs for special cases, showing that structures with larger distance between cycles exhibit a faster reduction rate.

Based on these results, we propose the PEG-CYCLE algorithm, which greedily maximizes the distance between cycles during the construction of the Tanner graph. This algorithm is applicable to both LDPC and QC-LDPC codes, including regular, irregular, fully connected, and non-fully connected cases. Simulations show that the codes constructed using our algorithm achieve performance comparable to or even better than those generated by state-of-the-art construction methods.

Our work introduces a novel approach for removing small ETSs in the Tanner graph by increasing the distance between cycles, thereby improving the performance of LDPC codes in the error floor region.

\subsection{Paper Outline}

The structure of this paper is as follows: Section~\ref{sec:definition} introduces essential definitions and notations. 
In Section~\ref{sec:theoretical}, we obtain theoretical results from two perspectives: the Tur\'an numbers of specific graphs and the spectral radius of the system matrix, and we explain their impact on ETSs. In Subsection~\ref{sec:theoretical turan}, we determine the Tur\'an numbers of specific dumbbell graphs and extend the results to general cases. In Subsection~\ref{sec:theoretical radius}, we use spectral theory to analyze the variation of spectral radius as the cycle length and distance between cycles change. Section~\ref{sec:6 cycles} examines the case of two 6-cycles with distances -1, 0, and 1, and calculates the sizes, spectral radii and the rate of error probability reduction of corresponding ETSs. Based on these findings, Section~\ref{sec:construction} designs the PEG-CYCLE algorithm and presents simulation results. Section~\ref{sec:conclusion} offers conclusion of this paper.

\section{Definitions and Preliminaries}\label{sec:definition}

In this paper, matrices are represented by bold capital letters (e.g., $\mathbf{A}$), and vectors by bold lowercase letters (e.g., $\boldsymbol{x}$). Their entries are denoted as $\mathbf{A}(i,j)$ for matrices and $\boldsymbol{x}(i)$ for vectors. Scalars or variables are represented by lowercase letters (e.g., $k$).

\subsection{Graph Theory}

We begin with some basic definitions in graph theory.
An undirected graph $G=(V(G),E(G))$ consists of a non-empty vertex set $V(G)$ and an edge set $E(G)$, where edges are unordered pairs of vertices.
Two vertices $u$ and $v$ in $V(G)$ are adjacent, or neighbors, if they are connected by an edge, denoted as $uv\in E(G)$. 
For an edge $uv\in E(G)$, we denote by $G_{uv}$ the graph obtained by subdividing the edge $uv$, which means introducing a new vertex on the edge $uv$.

The neighborhood of a vertex $v$ is the set of its neighbors and is denoted as $N_G(v)=\{u\in V(G)\mid uv\in E(G)\}$. The degree of a vertex $v\in G$, denoted as $d_G(v)$, is the total number of its neighbors. A graph $G$ is regular if all its vertices have the same degree. For brevity, we use $V$, $E$, $N(v)$, and $d(v)$ when no ambiguity arises. Let $\delta(G)$ denote the minimum degree of $G$. 
A graph without loops or multiple edges is a simple graph, which is the focus of this paper. 

A simple graph is complete, denoted $K_n$, if there is an edge connecting every pair of vertices in $G$. A simple graph is bipartite if its vertices can be partitioned into two sets $V_1$, $V_2$ such that no edge joins two vertices in the same set. Additionally, a bipartite graph is complete, if every vertex in $V_1$ is connected to every vertex in $V_2$, and is denoted $K_{m,n}$ with $m$ and $n$ vertices in each part, respectively.

Given a subset $S\subseteq V$, the neighborhoods of $S$ are denoted by $N(S)$. The induced subgraph generated by $S$ is defined as $G[S]=(S,E(G[S]))$, where $E(G[S])=\{uv\in E(G)\mid u,v \in S\}$.
We denote by $G- S$ the graph obtained from \( G \) by removing all vertices in \( S \) along with all edges incident to them.
For a subset $E'\subseteq E(G)$, $G-E'$ denotes the graph obtained by removing the edges in $E'$ from $G$, while keeping $V(G)$ unchanged. Specifically, for a vertex \( v \) and an edge \( e \), \( G - \{v\} \) and \( G - \{e\} \) are abbreviated as \( G - v \) and \( G - e \), respectively.

A path of length $k-1$, denoted by $P_k$, is a sequence of distinct vertices $v_1v_2\dots v_k$, where ${v_i}{v_{i+1}}\in E$ for all $1\leq i \leq k-1$. 
We say two or more paths are internally disjoint if they share no common vertices except the endpoints.
A graph $G$ is connected if, for any two distinct vertices $v_i$ and $v_j$, there exists a path from $v_i$ to $v_j$. A matching of size $k$ is a set of edges with no common vertices and is denoted as $k\cdot P_2$. 
A cycle of length $k$, denoted by $C_k$, is a path $P_k$ with an additional edge connecting the first and last vertices.

\begin{figure}[!tb]
    \centering
    \includegraphics[width=2.5in]{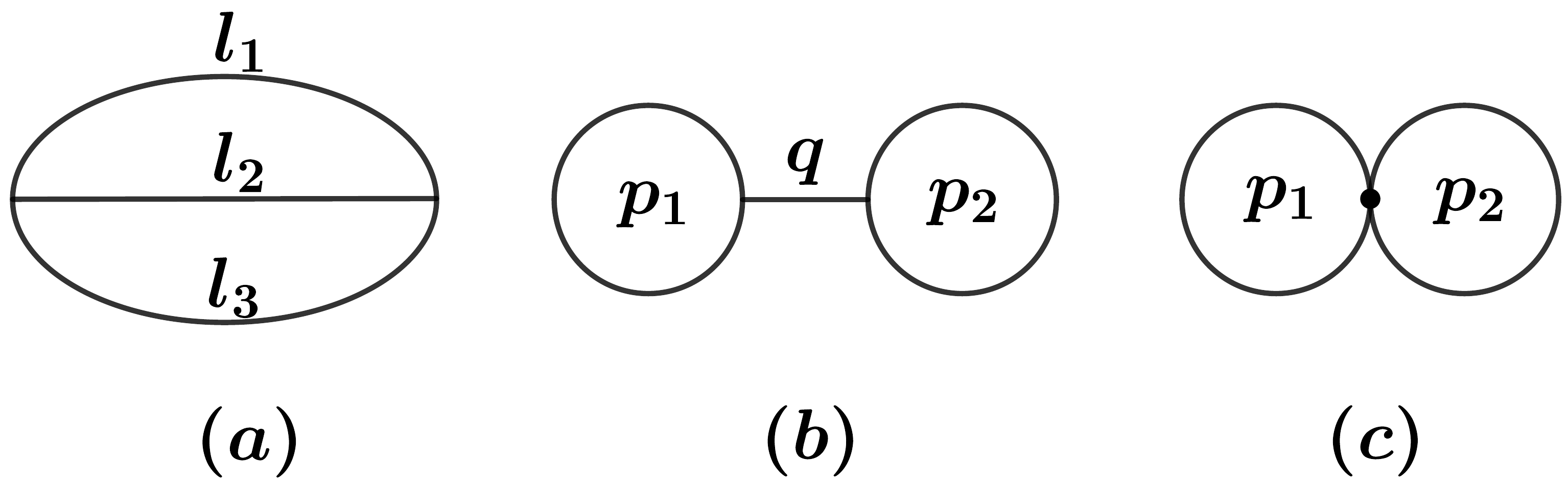}
    \caption{Figure $(a)$ is the theta graph $\theta(l_1,l_2,l_3)$. Figures $(b)$ and $(c)$ show the dumbbell graphs $D_b(r_1,r_2;q)$ for $q\geq 1$ and $q=0$, respectively.}
    \label{fig theta dumbbell}
\end{figure}

In this paper, we focus on the impact of distance between cycles on the performance of LDPC codes. To quantify the distance between two cycles, we introduce the following special graphs in graph theory, as shown in Fig.~\ref{fig theta dumbbell}.

\begin{definition}\label{def:theta}
    A theta graph $\theta(l_1,l_2,l_3)$ is the graph formed by three internally disjoint paths with common endpoints, where the three paths have lengths $l_1$, $l_2$, and $l_3$, respectively.
\end{definition}

\begin{definition}\label{def:dumbbell}
    A dumbbell graph $D_b(r_1,r_2;q)$ consists of two vertex-disjoint cycles, $C_{r_1}$ and $C_{r_2}$, with lengths $r_1$ and $r_2$, respectively, and a path $P_{q+1}$ of length $q$ ($q\geq 1$) with sharing only its endpoints with the two cycles.
    $D_b(r_1,r_2;0)$ represents that $C_{r_1}$ intersects $C_{r_2}$ at exactly one common vertex. 
\end{definition}

\begin{remark}
    To avoid repetitive discussions about isomorphic graphs, we assume that $l_1\leq l_2\leq l_3$ for the theta graph $\theta(l_1,l_2,l_3)$ and $r_1\leq r_2$ for the dumbbell graph $D_b(r_1,r_2;q)$. As there are no multiple edges, then $l_2\geq 2$ and $r_1$, $r_2\geq 3$.
\end{remark}

For the dumbbell graph $D_b(r_1,r_2;q)$, the distance between $C_{r_1}$ and $C_{r_2}$ can be naturally defined as $q$, the length of the path connecting them. In contrast, theta graphs are used to describe cases where two cycles share common edges, and their distance is defined as a negative number. Note that a theta graph $\theta(l_1,l_2,l_3)$ actually contains three cycles: $C_{l_1+l_2}$, $C_{l_1+l_3}$, and $C_{l_2+l_3}$. Thus, $\theta(l_1,l_2,l_3)$ can describe the following three cases:
\begin{itemize}
    \item $C_{l_1+l_2}$ and $C_{l_1+l_3}$ with a distance of $-l_1$;
    \item $C_{l_1+l_2}$ and $C_{l_2+l_3}$ with a distance of $-l_2$;
    \item $C_{l_1+l_3}$ and $C_{l_2+l_3}$ with a distance of $-l_3$.
\end{itemize}
To avoid redundancy and unnecessary discussion of isomorphic graphs, we measure the distance using the shortest path in a theta graph. Therefore, we interpret $\theta(l_1,l_2,l_3)$ as representing $C_{l_1+l_2}$ and $C_{l_1+l_3}$ with a distance of $-l_1$. 

Based on the above discussion, we define the distance between the cycles represented by the two graphs as follows:

\begin{definition}\label{def:distance}
    The dumbbell graph $D_b(r_1,r_2;q)$ represents a cycle of length $r_1$ and a cycle of length $r_2$ with distance $q$. The theta graph $\theta(l_1,l_2,l_3)$ represents a cycle of length $l_1+l_2$ and a cycle of length $l_1+l_3$ with distance $-l_1$.
\end{definition}

Since $l_1\leq l_2\leq l_3$, then $l_1\leq \frac{l_1+l_2}{2}$ and $l_1\leq \frac{l_1+l_3}{2}$. Therefore, for two cycles of lengths of $c_1$ and $c_2$, the distance $d$ satisfies $d\geq \max\{-\left\lfloor \frac{c_1}{2}\right\rfloor, -\left\lfloor \frac{c_2}{2}\right\rfloor\}$. 

Based on the above discussion, we can map different combinations of cycle lengths and distance between cycles to theta graphs or dumbbell graphs. To show the impact of the distance between cycles on ETSs, we introduce the following definitions from extremal graph theory.

\begin{definition}\label{def2}
    Let $H$ be a fixed graph. The Tur{\'a}n number $ex(n,{H})$ is the maximum number of edges in any graph of $n$ vertices that does not contain any subgraph isomorphic to ${H}$.
    An extremal graph of ${H}$ is an $n$-vertex graph with $ex(n,{H})$ edges and contains no subgraph isomorphic to ${H}$.
\end{definition}

The earliest result about Tur\'an numbers can be traced back to Mantel's theorem from 1907, which is stated as follows:

\begin{theorem}[Mantel's theorem,~\cite{mantel1907problem}]\label{thm:mantel}
    For all $n\geq 3$, $ex(n,C_3)=\left\lfloor \frac{n^2}{4}\right\rfloor$.
\end{theorem}

\subsection{Spectra of Graphs}

A directed graph (or digraph) $D = (V,A)$ consists of a set of vertices $V$ connected by directed edges, often called arcs, denoted by $A$. An arc $(u, v)$ is directed from $u$ (the head) to $v$ (the tail). 
A digraph is simple if it contains no loops or parallel arcs.

The adjacency matrix $\mathbf{A}(D)$ of a digraph $D$ is a $|V|\times |V|$ matrix, where the $(i,j)$ entry represents the number of arcs from $v_i$ to $v_j$. When $D$ is simple, $\mathbf{A}(D)$ is a $(0,1)$-matrix with all diagonal entries equal to 0. 

Let $\mathbf{M}$ be a real $n \times n$ matrix with nonnegative entries. $\mathbf{M}$ is called irreducible if for all $i, j$, there exists a positive integer $k$ such that
$(\mathbf{M}^k){(i,j)}>0$.
For a matrix $\mathbf{A}$, we write $\mathbf{A} > 0$ (or $\mathbf{A}\geq0$) to indicate that all entries of $\mathbf{A}$ are positive (nonnegative).
Similarly, for two matrices $\mathbf{A_1}$ and $\mathbf{A_2}$, we use $\mathbf{A_1}\geq \mathbf{A_2}$ to mean that $\mathbf{A_1}-\mathbf{A_2}\geq 0$.
The same notation applies to vectors.
The spectral radius $\rho(\mathbf{M})$ of $\mathbf{M}$ is defined as the maximum modulus of its eigenvalues:
\begin{equation*}
    \rho(\mathbf{M}) = \max \{|\lambda|: \lambda \text{ is an eigenvalue of }\mathbf{M}\}.
\end{equation*}

We introduce the following theorem about the eigenvalues of irreducible matrices, which is one of the most important results in matrix theory.

\begin{theorem}[Perron-Frobenius Theorem~\cite{brouwer2011spectra}]\label{thm:pfthm}
    Let $\mathbf{M}\geq 0$ be an $n\times n$ irreducible matrix. Then there exists a unique positive real number $\rho$ (the spectral radius of $\mathbf{M}$) with the following properties:
    \begin{itemize}
        \item[(i)] There exists a real vector $\boldsymbol{x_0}> 0$ such that $\mathbf{M}\boldsymbol{x_0}=\rho\boldsymbol{x_0}$.
        \item[(ii)] For every eigenvalue $\lambda $ of $\mathbf{M}$, $|\lambda|\leq \rho$.
        \item[(iii)] If $\boldsymbol{x}\geq 0 $, $\boldsymbol{x}\neq0$ and $\mathbf{M} \boldsymbol{x}\leq \rho'\boldsymbol{x}$, then $\boldsymbol{x} > 0$ and $\rho \leq \rho'$. Moreover, $\rho=\rho'$ if and only if $\mathbf{M} \boldsymbol{x} = \rho'\boldsymbol{x}$.
        \item[(iv)] The spectral radius $\rho$ satisfies the inequality:
        \begin{equation*}
            \min\limits_i\sum^n_{j=1}\mathbf{M}(i,j)\leq \rho\leq \max\limits_i\sum^n_{j=1}\mathbf{M}(i,j).
        \end{equation*} One of the equalities holds if and only if all row sums of $\mathbf{M}$ are equal, i.e., $\min\limits_i\sum^n_{j=1}\mathbf{M}(i,j)=\max\limits_i\sum^n_{j=1}\mathbf{M}(i,j)$.
    \end{itemize}
\end{theorem}

\subsection{Coding Theory}

A Tanner graph $G=(L \cup R,E)$ is a bipartite graph corresponding to the parity-check matrix $\mathbf{H}$ of an LDPC code. In this graph, the set $L$ represents variable nodes, each corresponding to a column in $\mathbf{H}$, while the set $R$ represents check nodes, each corresponding to a row in $\mathbf{H}$. An edge connects the $i$-th check node to the $j$-th variable node if the entry $\mathbf{H}(i,j)=1$. Specifically, the graph is variable-regular if every variable node has the same degree with $d_L(v)=\gamma$. 

For a fixed positive integer $p$, called the lifting degree, the parity-check matrix $\mathbf{H}$ can be represented by $p\times p$ circulant permutation matrices as follows~\cite{fossorier2004quasicyclic}:
\begin{equation}
    \label{eq1}
    \mathbf{H} = 
\begin{bmatrix}
    \mathbf{I}^{(p_{1,1})} & \mathbf{I}^{(p_{1,2})} & \cdots & \mathbf{I}^{(p_{1,\eta})}\\
    \mathbf{I}^{(p_{2,1})} & \mathbf{I}^{(p_{2,2})} & \cdots & \mathbf{I}^{(p_{2,\eta})}\\
    \vdots & \vdots & \ddots & \vdots\\
    \mathbf{I}^{(p_{\gamma,1})} & \mathbf{I}^{(p_{\gamma,2})} & \cdots & \mathbf{I}^{(p_{\gamma,\eta})}\\
\end{bmatrix}
\end{equation}

For $1\leq i\leq \gamma$ and $1\leq j\leq \eta$, $\mathbf{I}^{(p_{i,j})}$ represents a $p\times p$ circulant permutation matrix (CPM) with lifting value $p_{i,j}$, where $p_{i,j}\in \{0,1,2,\dots,p-1,\infty\}$. Specifically, $\mathbf{I}^{(\infty)}$ corresponds to a $p\times p$ zero matrix, while for $0\leq r\leq p-1$, $\mathbf{I}^{(p_{i,j})}$ has a `1' at position $(r,(r+p_{i,j}) \mod p)$ and `0's elsewhere, where $\mathbf{I}^{(0)}$ is the identity matrix. A QC-LDPC code is fully connected if its parity check matrix $H$ contains no $\mathbf{I}^{(\infty)}$ entry; otherwise, it is non-fully connected. Two auxiliary matrices are defined: the exponent matrix $\mathbf{E}=(p_{i,j})$ and the base matrix $\mathbf{B}=(b_{i,j})$, where $b_{i,j} = 1$ if $p_{i,j}\neq \infty$ and $0$ otherwise.

The necessary and sufficient condition for the existence of a length-$2k$ cycle in the Tanner graph can be characterized by the following equality~\cite{fossorier2004quasicyclic}:
\begin{equation}
    \sum_{i=0}^{k-1}(p_{m_i,n_i}-p_{m_i,n_{i+1}})\equiv 0 \mod p,\label{eq2}
\end{equation}
with $n_k = n_0$ and $m_i\neq m_{i+1}, n_i\neq n_{i+1}$ for all $0\leq i \leq k-1$:

For a subset $S\subseteq L$, an $(a,b)$ trapping set (TS) is an induced subgraph $G[S\cup N(S)]$ with $|S|=a$, and $b$ is the number of odd-degree vertices in $N(S)$. An elementary trapping set (ETS) is a special type of TS in which all the check nodes have degrees of either 1 or 2.

For a given ETS, a variable node (VN) graph $G_{VN}=(V_{VN},E_{VN})$ is constructed by first removing all check nodes of degree 1. The vertices $V_{VN}$ correspond to the variable nodes in the ETS, while the edges $E_{VN}$ correspond to the pairs of variable nodes connected through check nodes of degree 2~\cite{koetter2004pseudo}.

In a variable-regular Tanner graph with $d_L(v)=\gamma$, there exists a one-to-one correspondence between an $(a,b)$-ETS and its corresponding VN graph. The VN graph is obtained from the $(a,b)$-ETS by definition. Conversely, the $(a,b)$-ETS can be reconstructed from a given VN graph by placing a check node on each edge to represent check nodes of degree 2. Additionally, for every variable node $u\in V_{VN}$ with $d_{G_{VN}}(u)<\gamma$, we can attach $\gamma-d_{G_{VN}}(u)$ check nodes of degree 1, thereby obtaining the $(a,b)$-ETS. Therefore, in this paper, we often equate the VN graph with its corresponding ETS.

For a VN graph $G_{VN}=(V_{VN},E_{VN})$ of an $(a,b)$-ETS in a variable-regular Tanner graph with degree $d_L(v)=\gamma$, we have
$|V_{VN}|=a$ and $|E_{VN}|=\frac{1}{2}(a\gamma-b)$.
Since $|E_{VN}|$ is an integer, this implies:
\begin{itemize}
    \item when $\gamma$ is odd, $a$ and $b$ have the same parity;
    \item when $\gamma$ is even, $b$ must be even.
\end{itemize}
Moreover, if the girth of the Tanner graph exceeds 4, the VN graph of its ETS is simple.

We introduce the linear state-space model in~\cite{Butler2014errorfloor}, which characterizes the decoding behavior of an ETS.
This model is expressed by the following equations:
\begin{subequations}\label{eq:linear state-space model}
    \begin{align}
        &\boldsymbol{x}^{(0)} = \mathbf{B} \boldsymbol{\lambda};\label{model1}\\
        &\boldsymbol{x}^{(l)} = \mathbf{A_{sys}} \boldsymbol{x}^{(l-1)}+\mathbf{B} \boldsymbol{\lambda}+\mathbf{B_{ex}}\boldsymbol{\lambda_{ex}}^{(l)} \quad \text{ for $l\geq 1$};\label{model2}\\
        &\boldsymbol{\tilde{\lambda}}^{(l)} = \mathbf{C} \boldsymbol{x}^{(l-1)}+\boldsymbol{\lambda}+\mathbf{D_{ex}}\boldsymbol{\lambda_{ex}}^{(l)}\quad \text{ for $l\geq 1$}\label{model3},
    \end{align}
\end{subequations}
where: 
\begin{itemize}
    \item $\boldsymbol{x}^{(l)}$ represents the log-likelihood ratio (LLR) messages from variable nodes to degree-2 check nodes at iteration $l$.
    \item $\boldsymbol{\lambda}$ is the intrinsic information from the channel.
    \item $\boldsymbol{\lambda_{ex}}^{(l)}$ is the messages from the degree-1 check nodes at iteration $l$.
    \item $\boldsymbol{\tilde{\lambda}}^{(l)}$ represents the soft-output decisions (LLR) at iteration $l$.
\end{itemize}
For an $(a,b)$-ETS with $d_L(v)=\gamma$, there are $m=a\gamma-b$ edges connecting variable nodes and degree-2 check nodes. The vectors $\boldsymbol{\lambda}$ and $\boldsymbol{\tilde{\lambda}}^{(l)}$ have $a$ entries, while $\boldsymbol{\lambda_{ex}}^{(l)}$ has $b$ entries.

The system matrix $\mathbf{A_{sys}}$, an $m\times m$ matrix, describes how the messages on edges are updated between iterations. Each row and column of $\mathbf{A_{sys}}$ corresponds to an edge, with $\mathbf{A_{sys}}({i,j})=1$ indicating that the message on edge $j$ in the previous iteration contributes to the update of the message on edge $i$ in the current iteration. The diagonal elements of $\mathbf{A_{sys}}$ are `0's, and the sum of the entries in the $i$-th row equals $d(v_i)-1$, where $v_i$ is the head of edge $i$. Additionally, $\mathbf{A_{sys}}^{T}$ represents the adjacency matrix of the directed graph describing edge-message updates.

The matrices $\mathbf{B}$, $\mathbf{B_{ex}}$, $\mathbf{C}$, and $\mathbf{D_{ex}}$ describe relationships between channel information, messages from degree-1 check nodes, messages on the edges, and soft-output decisions. Specifically:
\begin{itemize}
    \item The matrix $\mathbf{B}$, of size $m\times a$, maps channel information to the messages on the edges.
    \item The matrix $\mathbf{B_{ex}}$, of size $m\times b$, maps messages from degree-1 check nodes to the messages on the edges.
    \item The matrix $\mathbf{C}$, of size $a\times m$, maps the messages on the edges to the soft-output decisions.
    \item The matrix $\mathbf{D_{ex}}$, of size $a\times b$, maps the messages from degree-1 check nodes to the soft-output decisions.
\end{itemize}

Equations~(\ref{model1})-(\ref{model3}) describe the message-passing process:
\begin{itemize}
    \item Equation~(\ref{model1}): In the initial stage, each variable node sends the intrinsic information from the channel to the degree-2 check nodes.
    \item Equation~(\ref{model2}): In iteration $l$, the messages transmitted from each variable node to degree-2 check nodes depend on three components: (1) the messages received from other degree-2 check nodes in the previous iteration, (2) the intrinsic channel information, and (3) the messages from degree-1 check nodes.
    \item Equation~(\ref{model3}): In iteration $l$, the soft-output decision at each variable node is the sum of three components: (1) the messages received from degree-2 check nodes, (2) the intrinsic channel information, and (3) the messages from degree-1 check nodes.
\end{itemize}
For a more detailed and complete discussion, readers can refer to~\cite{Butler2014errorfloor}.
We use the following example in \cite{Butler2014errorfloor} to further explain this model. 

\begin{figure}[!tb]
    \centering
    \includegraphics[width=2.5in]{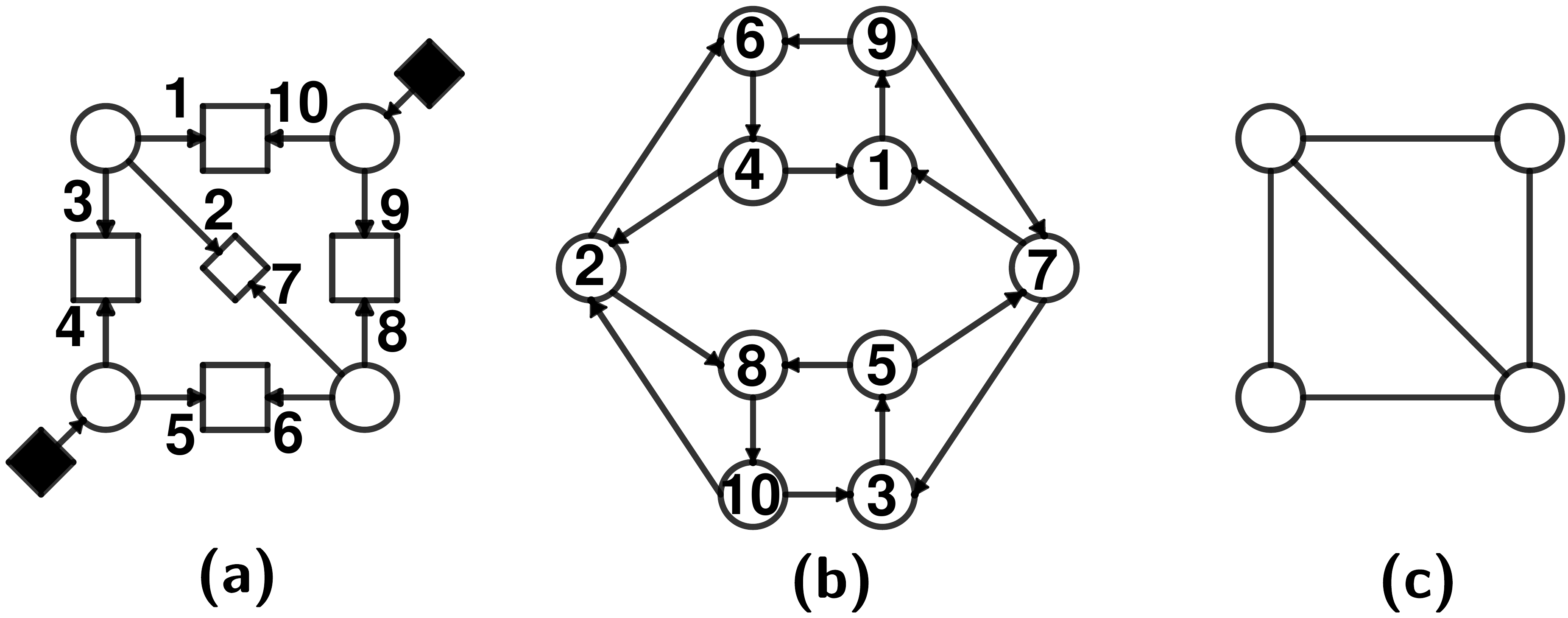}
    \caption{Figure $(a)$ labels each directed edge in the $(4,2)$-ETS, where circles represent variable nodes, squares represent check nodes, and black solid squares represent 1-degree check nodes. Figure $(b)$ illustrates the update relationships of the directed edges in this ETS, where each node corresponds to a directed edge from Figure $(a)$, and the directed edges indicate the message-passing process: the direction from the tail to the head signifies that the directed edge at the head participates in the update of the directed edge at the tail. Figure $(c)$ shows the corresponding VN graph.}
    \label{fig digraph}
\end{figure}

\begin{example}\label{ex:radius 42ets}
    Consider a $(4,2)$-ETS with $d_L(v)=3$. We begin by labeling each directed edge, as shown in Fig.~\ref{fig digraph}~$(a)$. Corresponding to this ETS, we provide explicit expressions for each matrix as follows:
    \begin{equation*}
        \mathbf{A_{sys}} = 
    \begin{bmatrix}
        0 & 0 & 0 & 1 & 0 & 0 & 1 & 0 & 0 & 0\\
        0 & 0 & 0 & 1 & 0 & 0 & 0 & 0 & 0 & 1\\
        0 & 0 & 0 & 0 & 0 & 0 & 1 & 0 & 0 & 1\\
        0 & 0 & 0 & 0 & 0 & 1 & 0 & 0 & 0 & 0\\
        0 & 0 & 1 & 0 & 0 & 0 & 0 & 0 & 0 & 0\\
        0 & 1 & 0 & 0 & 0 & 0 & 0 & 0 & 1 & 0\\
        0 & 0 & 0 & 0 & 1 & 0 & 0 & 0 & 1 & 0\\
        0 & 1 & 0 & 0 & 1 & 0 & 0 & 0 & 0 & 0\\
        1 & 0 & 0 & 0 & 0 & 0 & 0 & 0 & 0 & 0\\
        0 & 0 & 0 & 0 & 0 & 0 & 0 & 1 & 0 & 0\\
    \end{bmatrix}
    \end{equation*}

    \begin{equation*}
        \mathbf{B} = 
    \begin{bmatrix}
        1 & 0 & 0 & 0\\
        1 & 0 & 0 & 0\\
        1 & 0 & 0 & 0\\
        0 & 1 & 0 & 0\\
        0 & 1 & 0 & 0\\
        0 & 0 & 1 & 0\\
        0 & 0 & 1 & 0\\
        0 & 0 & 1 & 0\\
        0 & 0 & 0 & 1\\
        0 & 0 & 0 & 1\\
    \end{bmatrix}
    \mathbf{B_{ex}} = 
    \begin{bmatrix}
        0 & 0\\
        0 & 0\\
        0 & 0\\
        1 & 0\\
        1 & 0\\
        0 & 0\\
        0 & 0\\
        0 & 0\\
        0 & 1\\
        0 & 1\\
    \end{bmatrix}
    \end{equation*}

    \begin{equation*}
        \mathbf{C} = 
    \begin{bmatrix}
        0 & 0 & 0 & 1 & 0 & 0 & 1 & 0 & 0 & 1\\
        0 & 0 & 1 & 0 & 0 & 1 & 0 & 0 & 0 & 0\\
        0 & 1 & 0 & 0 & 1 & 0 & 0 & 0 & 1 & 0\\
        1 & 0 & 0 & 0 & 0 & 0 & 0 & 1 & 0 & 0\\
    \end{bmatrix}
    \mathbf{D_{ex}} = 
    \begin{bmatrix}
        0 & 0\\
        1 & 0\\
        0 & 0\\
        0 & 1\\
    \end{bmatrix}.
    \end{equation*}
    Note that $\mathbf{A_{sys}^T}$ corresponds to the adjacency matrix of Fig.~\ref{fig digraph}~$(b)$. Specifically, $\mathbf{A_{sys}}(i,j)=1$ indicates that the edge labeled $j$ participates in the update of the edge labeled $i$. Therefore, there is a directed edge from $j$ to $i$ in the directed graph, as shown in Fig.~\ref{fig digraph}~$(b)$. For example, the first row of $\mathbf{A_{sys}}$ indicates that the information of edge $1$ is updated by the information from edge 4 and 7. This corresponds to two directed edges, $(4,1)$ and $(7,1)$, in the digraph shown in Fig.~\ref{fig digraph}~$(b)$.
\end{example}

In this linear state-space model, the spectral radius of $\mathbf{A_{sys}}$ of the ETS is an important parameter that influences the decoding behavior. As shown in~\cite{Butler2014errorfloor}, the authors have shown that when errors occur in the ETS, a larger spectral radius leads to slower error correction, and in some cases, it may even result in incorrect decoding. Therefore, we regard the spectral radius of $\mathbf{A_{sys}}$ as a key indicator of the harmfulness of the ETS, with a larger spectral radius signifying a more harmful effect.

\begin{table*}[!tb]
    \begin{center}
    \caption{Summary of parameters and notations.}
    \label{tab:notations}
    \begin{tabular}{|c|c|}
    \hline
    $G=(V,E)$ & an undirected graph $G$ with vertex set $V$ and edge set $E$ \\ \hline
    $D=(V,A)$ & a directed graph $D$ with vertex set $V$ and arc set $A$ \\ \hline
    $G_{uv}$ & the graph obtained from $G$ by subdividing the edge $uv$ \\ \hline
    $d(v)$ & degree of a vertex $v$ \\ \hline
    $\delta(G)$ & the minimum degree of $G$ \\ \hline
    $G[S]$ & the induced subgraph generated by $S\subseteq V$ \\ \hline
    $K_n$ & a complete graph with $n$ vertices \\ \hline
    $K_{m,n}$ & a complete bipartite graph with $m$ and $n$ vertices in each part \\ \hline
    $P_k$ & a path of length $k-1$ \\ \hline
    $k\cdot P_2$ & a matching of size $k$ \\ \hline
    $C_k$ & a cycle of length $k$ \\ \hline
    $\theta(l_1,l_2,l_3)$ & a theta graph where the three paths are of lengths $l_1$, $l_2$, and $l_3$ \\ \hline
    $D_b(r_1,r_2;q)$ & a dumbbell graph consisting of two vertex-disjoint cycles, $C_{r_1}$ and $C_{r_2}$, and a path $P_{q+1}$ \\ \hline
    $ex(n,H)$ & Tur\'an number of $H$ \\ \hline
    $\mathbf{A}(D)$ & the adjacency matrix of a digraph $D$ \\ \hline
    $\rho(\mathbf{M})$ & the spectral radius of a matrix $\mathbf{M}$ \\ \hline
    $d_L(v)$ & the degree of variable nodes in a variable-regular Tanner graph \\ \hline
    $g$ & the girth of Tanner graph \\ \hline
    $p$ & the lifting degree of a QC-LDPC code \\ \hline
    $\mathbf{I}^{(p_{i,j})}$ & a circulant permutation matrix with lifting value $p_{i,j}$ \\ \hline
    $\mathbf{A_{sys}}$ & the system matrix of an ETS \\ \hline
    \end{tabular}
    \end{center}
\end{table*}

For the reader's convenience, the parameters used in this paper are summarized in Table~\ref{tab:notations}.

\section{Theoretical Results}\label{sec:theoretical}

In this section, we obtain results on Tur\'an numbers and the spectral radius of the system matrix for various cycle lengths and different distances between cycles. Our theoretical analysis consistently shows that longer cycles and larger distances between cycles in the Tanner graphs lead to improved performance of the corresponding LDPC codes in the error floor region.

\subsection{Tur\'an Numbers of Two Classes of Graphs}\label{sec:theoretical turan}

We first determine the Tur\'an numbers for the dumbbell graphs $D_b(3,3;0)$ and $D_b(3,3;1)$, in comparison with the result for the Tur\'an number of $\theta(1,2,2)$ in~\cite{xiong2024theoretical}:

\begin{theorem}[\cite{xiong2024theoretical}]\label{thm:t122}
    For all $n\geq4$, $ex(n,\theta(1,2,2))=\lfloor \frac{n^2}{4}\rfloor$.
\end{theorem}

We begin by proving the following lemma, which is necessary for determining the Tur\'an numbers of the two dumbbell graphs.

\begin{lemma}\label{lemma:induction}
    Let $n$ be a positive integer and $f: \mathbb{N} \rightarrow \mathbb{N} $ be a function satisfying the inequality $(n-2)(f(n)+1)-nf(n-1)>0$.
    For a fixed graph $H$, if $ex(n-1,H)\leq f(n-1)$, then $ex(n,H)\leq f(n)$.
\end{lemma}

\begin{IEEEproof}
    Assume that $G$ is a graph with $n$ vertices and $f(n)+1$ edges.
    We aim to prove that $G$ must contain a copy of $H$. 
    First, suppose that no vertex $v_0$ in $G$ has degree $d(v_0) \leq f(n)-f(n-1)$. Then, for every vertex $v$ in $G$, we have $d(v) \geq f(n)-f(n-1)+1$. Consequently, the total number of edges satisfies 
    \begin{equation*}
        |E(G)|=\frac{1}{2}\sum_{v\in V(G)}d(v)\geq \frac{1}{2}n(f(n)-f(n-1)+1).
    \end{equation*}
    On the other hand, we know that $(n-2)(f(n)+1)-nf(n-1)>0$, which implies $f(n-1)<\frac{n-2}{n}(f(n)+1)$.
    Therefore, we obtain
    \begin{eqnarray*}
        |E(G)|&\geq& \frac{1}{2}n(f(n)-f(n-1)+1)\\
        &>&\frac{1}{2}n(f(n)-\frac{n-2}{n}(f(n)+1)+1)\\
        &=&f(n)+1,
    \end{eqnarray*}
    However, this contradicts the assumption that $|E(G)|=f(n)+1$.
    Therefore, there must exist a vertex $v_0$ with degree $d(v_0)\leq f(n)-f(n-1)$. 

    Next, consider the graph $G-v_0$, where $|V(G-v_0)| = n-1$ and $|E(G-v_0)| \geq f(n-1) + 1$. Since by assumption $ex(n-1,H) \leq f(n-1)$, it follows that $G-v_0$, and thus $G$, must contain a copy of $H$.
\end{IEEEproof}

The following corollary can be easily proven by induction.

\begin{corollary}\label{cor:induction}
    In particular, if there exists $n_0\in \mathbb{N}$ such that $ex(n_0,H)\leq f(n_0)$ and the inequality $(n-2)(f(n)+1)-nf(n-1)>0$ holds for all $n\geq n_0+1$, then for any $n\geq n_0$, we have $ex(n,H)\leq f(n)$.
\end{corollary}

Based on Lemma~\ref{lemma:induction} and Corollary~\ref{cor:induction}, we can determine the exact values of $ex(n,D_b(3,3;0))$ and $ex(n,D_b(3,3;1))$.

\begin{theorem}\label{thm:d330}
    For all $n\geq5$, $ex(n,D_b(3,3;0))=\lfloor \frac{n^2}{4}\rfloor+1$.
\end{theorem}
    
\begin{IEEEproof}
    For the lower bound, consider a graph $G_0$ on $n$ vertices with $E(G_0)=E(K_{\lfloor \frac{n}{2}\rfloor,\lceil \frac{n}{2}\rceil})\cup \{xy\}$, where $x$ and $y$ are vertices not adjacent in $K_{\lfloor \frac{n}{2}\rfloor,\lceil \frac{n}{2}\rceil}$. It can be easily verified that $G_0$ does not contain a copy of $D_b(3,3;0)$, and thus, 
    $ex(n,D_b(3,3;0))\geq \lfloor \frac{n^2}{4}\rfloor+1$.
    For the upper bound, let $f(n) = \lfloor \frac{n^2}{4}\rfloor+1$. We now prove that $ex(n,D_b(3,3;0))\leq f(n)$.

    For $n=5$, consider a graph with $\lfloor \frac{n^2}{4}\rfloor+2=8$ edges. Then it is either $K_5-E(P_3)$ or $K_5-E(2\cdot P_2)$, both of which contain a copy of $D_b(3,3;0)$.

    For $n=6$, the inequality $(n-2)(f(n)+1)-nf(n-1)>0$ holds, implying $ex(6,D_b(3,3;0))\leq \lfloor \frac{6^2}{4}\rfloor+1=10$.

    For $n=7$, consider any graph $G$ with $|E(G)|=f(n)+1=14$. 
    If there exists a vertex $v'$ in $G$ such that $d(v')\leq 3$, then removing $v'$ leaves $G-v'$ with 6 vertices and at least 11 edges. Given $ex(6,D_b(3,3;0))=10$, $G$ must contain a copy of $D_b(3,3;0)$. If not, each vertex $v$ in $G$ has degree $d(v)\geq 4$, then with $|V(G)|=7$ and $|E(G)|=14$, $G$ is 4-regular.
    Choose a vertex $v_0$ and denote its neighbors as $\{v_1,v_2,v_3,v_4\}$, with the remaining vertices being $v_5$ and $v_6$. Since $d(v_1)=4$, then $v_1$ must be adjacent to at least one vertex in $\{v_2, v_3, v_4\}$, saying $v_2$. In the induced subgraph $G[\{v_3,v_4,v_5,v_6\}]$, there are at least $14-4\times3+3=5$ edges. By Theorem~\ref{thm:mantel}, there must be a $C_3$ in $G[\{v_3,v_4,v_5,v_6\}]$. If this $C_3$ is $\{v_3,v_4,v_5\}$ or $\{v_3,v_4,v_6\}$, then the induced subgraph of $\{v_0,v_1,v_2,v_3,v_4\}$ contains $D_b(3,3;0)$. If the $C_3$ is $\{v_3,v_5,v_6\}$ (or similarly $\{v_4,v_5,v_6\}$), since $d(v_3)=4$, another neighbor of $v_3$, saying $v_i$, will result in induced subgraph of $\{v_0,v_i,v_3,v_5,v_6\}$ containing $D_b(3,3;0)$.

    For $n\geq 8$:
    \begin{itemize}
        \item If $n=2k$, $k\geq 4$,
        \begin{eqnarray*}
            &&(n-2)(f(n)+1)-nf(n-1)\\
            &=&(2k-2)(k^2+2)-2k(k^2-k+1)\\
            &=&2k-4>0.
        \end{eqnarray*}
        \item If $n=2k+1$, $k\geq 4$,
        \begin{eqnarray*}
            &&(n-2)(f(n)+1)-nf(n-1)\\
            &=&(2k-1)(k^2+k+2)-(2k+1)(k^2+1)\\
            &=&k-3>0.
        \end{eqnarray*}
    \end{itemize}
    Thus, by Corollary~\ref{cor:induction}, we have $ex(n,D_b(3,3;0))\leq \lfloor \frac{n^2}{4}\rfloor+1$ for all $n\geq 8$. 
    Combining the results for $n=5$, $n=6$, and $n=7$, the proof is complete.
\end{IEEEproof}

\begin{theorem}\label{thm:d331}
    \begin{equation*}
        ex(n,D_b(3,3;1))=\left\{
        \begin{aligned}
            &12, \quad \text{if $n=6;$} \\
            &\lfloor \frac{n^2}{4}\rfloor+\lceil \frac{n}{2}\rceil-1, \quad \text{if $n\geq 7.$} 
        \end{aligned}
        \right.
    \end{equation*}
\end{theorem}
        
\begin{IEEEproof}
    If $n=6$, the complete bipartite graph $K_{3,3}$ with three additional edges $\{xy,yz,xz\}$ added within the same partition does not contain a $D_b(3,3;1)$. However, both $K_6-E(P_3)$ and $K_6-E(2\cdot P_2)$ contain a $D_b(3,3;1)$. Therefore, $ex(6,D_b(3,3;1))=12$.

    If $n\geq 7$, for the lower bound, consider a graph $G_0$ with $E(G_0)=E(K_{\lfloor \frac{n}{2}\rfloor,\lceil \frac{n}{2}\rceil}) \cup E(K_{1,\lceil \frac{n}{2}\rceil-1})$, where $K_{1,\lceil \frac{n}{2}\rceil-1}$ is in the same part of $K_{\lfloor \frac{n}{2}\rfloor,\lceil \frac{n}{2}\rceil}$. One can verify that there is no $D_b(3,3;1)$ in $G_0$, 
    so we conclude that $ex(n,D_b(3,3;1))\geq \lfloor \frac{n^2}{4}\rfloor+\lceil \frac{n}{2}\rceil-1$. Now, we will prove $ex(n,D_b(3,3;1))\leq \lfloor \frac{n^2}{4}\rfloor+\lceil \frac{n}{2}\rceil-1$ for all $n\geq7$.

    For $n=7$, consider a graph $G_1$ with 16 edges.
    If there exists a vertex $t$ with degree at most 3, then $G_1-t$ has 6 vertices and more than 12 edges, which guarantees the presence of a $D_b(3,3;1)$. If each vertex in $G_1$ has a degree of at least 4, the possible degree sequences are $\{6,6,4,4,4,4,4\}$ or $\{6,5,5,4,4,4,4\}$ or $\{5,5,5,5,4,4,4\}$.
    For these three degree sequences, we denote the corresponding vertices as $\{t_1,t_2,\dots,t_7\}$.
    If the degree sequence is $\{6,6,4,4,4,4,4\}$ or $\{6,5,5,4,4,4,4\}$, by Pigeonhole principle, there exist two adjacent vertices $t_i$ and $t_j$ with $d(t_i)=d(t_j)=4$. Then $t_i$, $t_j$ and $t_1$ form a $C_3$, and by Theorem~\ref{thm:mantel}, there is a $C_3$ in $G_1-\{t_1,t_i,t_j\}$. The two $C_3$ must be connected by an edge incident to $t_1$ as $d(t_1)=6$, thus forming a $D_b(3,3;1)$. If the degree sequence is $\{5,5,5,5,4,4,4\}$, similarly, by Pigeonhole principle, there is a $C_3$ with vertices of degrees 5, 5 and 4, and the induced graph by the remaining vertices contains a $C_3$ by Theorem~\ref{thm:mantel}. Hence, we obtain a $D_b(3,3;1)$ and conclude that $ex(7,D_b(3,3;1))\leq 15$.

    For $n\geq 8$, define $f(n)=\lfloor \frac{n^2}{4}\rfloor+\lceil \frac{n}{2}\rceil-1$. For even $n=2k$, consider a graph $G=(V,E)$ with $|V|=2k$ and $|E|=f(2k)+1=k^2+k$. If there exists a vertex $v_0$ with degree $d(v_0)\leq k$, then for the graph $G-v_0$ (with $2k-1$ vertices), the number of edges satisfies $|E(G-v_0)|\geq k^2=f(2k-1)+1$, implying the presence of a $D_b(3,3;1)$ in $G$. Otherwise, if each vertex has degree at least $k+1$, then $G$ is $(k+1)$-regular, since $|E|=k^2+k$. We claim that there is a $D_b(3,3;1)$ in this case, which, by assumption that $ex(2k-1,D_b(3,3;1))\leq f(2k-1)$, implies $ex(2k,D_b(3,3;1))\leq f(2k)$. 
    
    Fix a vertex $v_0$ in $G$ with neighborhood $N(v_0)=\{v_1,v_2,\dots,v_{k+1}\}$ and denote the remaining $k-2$ vertices as $\{u_1,u_2,\dots,u_{k-2}\}$. Note that $v_1$ must be adjacent to at least one vertex in $N(v_0)$, saying $v_2$.
    Consider the subgraph $G'=G-\{v_0,v_1,v_2\}$, which contains $2k-3$ vertices and
    $E(G')=k(k+1)-3(k+1)+3=k^2-2k\geq \lfloor \frac{(2k-3)^2}{4}\rfloor+1$.
    By Theorem~\ref{thm:mantel}, there exists a $C_3$ in $G'$. 
    Let $\{w_1,w_2,w_3\}$ be the vertices of this $C_3$. If any vertex in $\{w_1,w_2,w_3\}$ is adjacent to $v_0$, then the induced subgraph by $\{v_0,v_1,v_2,w_1,w_2,w_3\}$ contains a copy of $D_b(3,3;1)$. If none of $w_1$, $w_2$, $w_3$ is adjacent to $v_0$, then $\{w_1,w_2,w_3\}\subseteq \{u_1,u_2,\dots,u_{k-2}\}$, and as $d(w_1)=k+1$, $w_1$ must be adjacent to at least one vertex in $N(v_0)$. Suppose it is $v_i$.
    Similarly, $v_i$ must be adjacent to some vertex $v_j$ in $N(v_0)$ as $d(v_i)=k+1$. Hence, the induced subgraph by $\{v_0,v_i,v_j,w_1,w_2,w_3\}$ contains a copy of $D_b(3,3;1)$.
    Thus, there exists a $D_b(3,3;1)$ in $G$.

    If $n=2k+1$, $k\geq 4$, we have $(n-2)(f(n)+1)-nf(n-1)=(2k-1)(\lfloor \frac{(2k+1)^2}{4}\rfloor+\lceil \frac{2k+1}{2}\rceil)-(2k+1)(\lfloor \frac{(2k)^2}{4}\rfloor+\lceil \frac{2k}{2}\rceil-1)$, which simplifies to $(2k-1)(k^2+2k+1)-(2k+1)(k^2+k-1)=k>0$.
    By Lemma~\ref{lemma:induction}, $ex(2k,D_b(3,3;1))\leq f(2k)$ implies $ex(2k+1,D_b(3,3;1))\leq f(2k+1)$.
    By induction, the proof is complete.
\end{IEEEproof}

For general dumbbell graphs with at least one odd cycle, we have:

\begin{theorem}\label{thm:odd_even}
    Let $r_1$, $r_2$, and $q$ be positive integers, where $r_1\geq 3$, $r_2\geq 4$, $q\geq0$, and $r_1+r_2$ is odd.
    For $n\geq k^2+k$ with $k=3r_1+3r_2+2q-12$, then $ex(n,D_b(r_1,r_2;q))=\lfloor \frac{n^2}{4}\rfloor$.
\end{theorem}

\begin{IEEEproof}
    The proof is given in Appendix~\ref{appendix A}.
\end{IEEEproof}

\begin{theorem}\label{thm:odd_odd_1}
    Let $r_1$, $r_2$, and $q$ be positive integers, with $r_1$, $r_2$ odd, $3\leq r_1\leq r_2$, and $q\geq0$. For $n\geq k^2+k$ with $k=3r_1+3r_2+2q-11$, then
    \begin{equation*}
                ex(n,D_b(r_1,r_2;q))=\left\{
                \begin{aligned}
                    &\lfloor \frac{n^2}{4}\rfloor+1,\quad \text{when $q=0$;}\\
                    &\lfloor \frac{n^2}{4}\rfloor+\lceil \frac{n}{2}\rceil-1, \quad \text{when $q\geq1$.} 
                \end{aligned}
                \right.
            \end{equation*}
\end{theorem}

\begin{IEEEproof}
    The proof is given in Appendix~\ref{appendix B}.
\end{IEEEproof}

In~\cite{zhai2021turan}, the authors obtained the exact value of $ex(n,\theta(l_1,l_2,l_3))$ with at least one odd cycle.

\begin{theorem}[{\cite{zhai2021turan}}]
    Let $l_1$, $l_2$, and $l_3$ be three positive integers with different parities, and assume $l_1\leq l_2\leq l_3$. For $n\geq 9k^2-3k$ with $k=l_1+l_2+l_3-1$, then $ex(n,\theta(l_1,l_2,l_3))=\lfloor \frac{n^2}{4}\rfloor$.
\end{theorem}

\begin{remark}\label{remark:turan ets}
    In a Tanner graph with variable-regular degree $d_L(v)=\gamma$, the VN graph of an $(a,b)$-ETS has $a$ vertices and $\frac{1}{2}(a\gamma-b)$ edges. If $H$ is forbidden in the VN graph, by the definition of the Tur\'an number, we have $\frac{1}{2}(a\gamma-b)\leq ex(a,H)$, leading to the inequality $b\geq a\gamma-2ex(a,H)$. Consequently, all $(a,b)$-ETSs with $b< a\gamma-2ex(a,H)$ are eliminated. Therefore, for two fixed graphs $H_1$ and $H_2$ and a fixed positive integer $a$, if $ex(a,H_1)< ex(a,H_2)$, then $a\gamma-2ex(a,H_1)> a\gamma-2ex(a,H_2)$. This implies that removing $H_1$ from the VN graph can eliminate more $(a,b)$-ETSs than removing $H_2$.
\end{remark}

For LDPC codes, $(a,b)$-ETSs with smaller values of $a$ and $b$ are more harmful to the performance in the error floor region. Additionally, by Remark~\ref{remark:turan ets}, removing structures with smaller Tur\'an numbers can eliminate more small ETSs, making it more effective for improving the performance of LDPC codes in the error floor region. Furthermore, we observe that the Tur\'an numbers of dumbbell graphs are generally greater than those of theta graphs. According to Definition~\ref{def:distance}, this implies that structures with smaller distances between cycles in the Tanner graph are more harmful.

In the next subsection, we analyze the spectral radius of the system matrix $A_{sys}$ in the linear state-space model.

\subsection{Spectral Radius}\label{sec:theoretical radius}

We first prove the theorem regarding the trend of the spectral radius of the system matrix when subdividing an edge in a graph. This result suggests that for ETSs, their harmful effect decreases as the cycle length or the distance between cycles increases. Subsequently, we calculate the spectral radii of the system matrices for several representative theta graphs and dumbbell graphs with varying cycle lengths and distances. The numerical results are consistent with our theoretical conclusions.

For a simple graph $G$, let $D(G) = \{(u,v),(v,u)\mid uv \in E(G)\}$. Regarding $G$ as the VN graph of an ETS, we can define the system matrix $\mathbf{A_{sys}}(G)$ in two ways:
\begin{itemize}
    \item Reconstruct the ETS from $G$ and obtain $\mathbf{A_{sys}}(G)$ based on the ETS.
    \item Directly define $\mathbf{A_{sys}}(G)$ as a $|D(G)|\times |D(G)|$ matrix, where each row and column corresponds to an ordered pair in $D(G)$. Here, $\mathbf{A_{sys}}(G)(i,j)=1$ if the second entry of $j$ equals the first entry of $i$, and the first entry of $j$ differs from the second entry of $i$; otherwise, $\mathbf{A_{sys}}(G)(i,j)=0$.
\end{itemize}
Note that the two ways are equivalent. For example, the matrix $\mathbf{A_{sys}}$ in Example~\ref{ex:radius 42ets} is the system matrix of the VN graph shown in Fig.~\ref{fig digraph}~$(c)$.

\begin{lemma}\label{lemma:irreducible}
    Let $G$ be a connected graph without any vertex of degree one. If $G$ is not a cycle, then the system matrix $\mathbf{A_{sys}}(G)$ is irreducible.
\end{lemma}

\begin{IEEEproof}
    See~\cite[Theorem 6]{Butler2014errorfloor}.
\end{IEEEproof}

\begin{theorem}\label{thm:add edge}
    Let $G$ be a connected graph without any vertex of degree one, and suppose $G$ is not a cycle. For an edge $uv\in E(G)$, denote the system matrix of $G$ and $G_{uv}$ as $\mathbf{A_{sys}}(G)$ and $\mathbf{A_{sys}}(G_{uv})$, respectively. Then we have $\rho(\mathbf{A_{sys}}(G_{uv}))\leq \rho(\mathbf{A_{sys}}(G))$.
\end{theorem}

\begin{IEEEproof}
    By Lemma~\ref{lemma:irreducible}, both $\mathbf{A_{sys}}(G)$ and $\mathbf{A_{sys}}(G_{uv})$ are irreducible. We now show that there exists a vector $\boldsymbol{x'}\geq 0$ and $\boldsymbol{x'}\neq 0$ such that $\mathbf{A_{sys}}(G_{uv})\boldsymbol{x'}\leq \rho(\mathbf{A_{sys}}(G))\boldsymbol{x'}$. By Theorem~\ref{thm:pfthm} (iii), we can conclude $\rho(\mathbf{A_{sys}}(G_{uv}))\leq \rho(\mathbf{A_{sys}}(G))$. 

    Assume the eigenvector of $\mathbf{A_{sys}}(G)$ corresponding to $\rho(\mathbf{A_{sys}}(G))$ is $\boldsymbol{x}$, i.e., $\mathbf{A_{sys}}(G)\boldsymbol{x}=\rho (\mathbf{A_{sys}}(G))\boldsymbol{x}$ and $\boldsymbol{x}\geq 0$. Denote the new vertex on the edge $uv$ as $w$. We construct the vector $\boldsymbol{x'}$ as $\boldsymbol{x'}({(u,w)})=\boldsymbol{x'}({(w,v)})=\boldsymbol{x}({(u,v)})$, $\boldsymbol{x'}({(v,w)}) = \boldsymbol{x'}({(w,u)})=\boldsymbol{x}({(v,u)})$, and $\boldsymbol{x'}({a}) = \boldsymbol{x}({a})$ for other arcs $a\neq (u,w)$, $(w,v)$, $(v,w)$, and $(w,u)$. 
  
    Denote $\boldsymbol{y'}=\mathbf{A_{sys}}(G_{uv})\boldsymbol{x'}$. We prove that $\boldsymbol{y'}\leq \rho(\mathbf{A_{sys}}(G))\boldsymbol{x'}$ by examining three cases:
    \begin{itemize}
        \item For any arc $(v_i,v_j)$ with $v_i\neq u, v, w$, 
        \begin{eqnarray*}
            \boldsymbol{y'}({(v_i,v_j)})&=&\sum_{(v_k,v_i)\in D(G_{uv}),v_k\neq v_j}\boldsymbol{x'}({(v_k,v_i)})\\
            &=&\sum_{(v_k,v_i)\in D(G),v_k\neq v_j}\boldsymbol{x}({(v_k,v_i)})\\
            &=&(\mathbf{A_{sys}}(G)\boldsymbol{x})({(v_i,v_j)})\\
            &=&\rho (\mathbf{A_{sys}}(G))\boldsymbol{x'}({(v_i,v_j)}).
        \end{eqnarray*}
        \item For any arc $(u,v_j)$ with $v_j\neq w,v$,
        \begin{eqnarray*}
            \boldsymbol{y'}({(u,v_j)})
            &=&\sum_{(v_k,u)\in D(G_{uv}),v_k\neq v_j,w}\boldsymbol{x'}({(v_k,u)})+\boldsymbol{x'}({(w,u)})\\
            &=&\sum_{(v_k,u)\in D(G),v_k\neq v_j,v}\boldsymbol{x}({(v_k,u)})+\boldsymbol{x}({(v,u)})\\
            &=&(\mathbf{A_{sys}}(G)\boldsymbol{x})({(u,v_j)})\\
            &=&\rho (\mathbf{A_{sys}}(G))\boldsymbol{x'}({(u,v_j)}).
        \end{eqnarray*}
        Similarly, we can obtain $\boldsymbol{y'}({(v,v_j)})=\rho (\mathbf{A_{sys}}(G))\boldsymbol{x'}({(v,v_j)})$ for any arc $(v,v_j)$ with $v_j\neq u,w$.
        \item For the remaining four arcs $(u,w)$, $(w,u)$, $(w,v)$, and $(v,w)$, we have
        \begin{eqnarray*}
            \boldsymbol{y'}({(u,w)})&=&\sum_{(v_k,u)\in D(G_{uv}),v_k\neq w,v}\boldsymbol{x'}({(v_k,u)})\\
            &=&\sum_{(v_k,u)\in D(G),v_k\neq v}\boldsymbol{x}({(v_k,u)})\\
            &=&(\mathbf{A_{sys}}(G)\boldsymbol{x})({(u,v)})\\
            &=&\rho (\mathbf{A_{sys}}(G))\boldsymbol{x'}({(u,w)}).
        \end{eqnarray*}
        As there are no degree-one vertices in $G$, by Theorem~\ref{thm:pfthm} (iv), $\rho(\mathbf{A_{sys}}(G))\geq 1$. Then we have
        \begin{eqnarray*}
            \boldsymbol{y'}_{(w,u)}&=&(\mathbf{A_{sys}}(G_{uv})\boldsymbol{x'})({(w,u)}) =\boldsymbol{x'}({(v,w)})\\
            &=&\boldsymbol{x'}({(w,u)})
            \leq \rho (\mathbf{A_{sys}}(G))\boldsymbol{x'}({(w,u)}).
        \end{eqnarray*}
        Similarly,
        $\boldsymbol{y'}({(v,w)})=\rho (\mathbf{A_{sys}}(G))\boldsymbol{x'}({(v,w)})$, $\boldsymbol{y'}({(w,v)})\leq  \rho (\mathbf{A_{sys}}(G))\boldsymbol{x'}({(w,v)})$.
    \end{itemize}
    Then we have $\boldsymbol{y'}\leq \rho(\mathbf{A_{sys}}(G))\boldsymbol{x'}$ and finish the proof.
\end{IEEEproof}

\begin{corollary}\label{cor:radius}
    Let $G$ be a connected graph with no vertex of degree one, and assume that $G$ is not a cycle. If $G$ is not regular, then for an edge $uv\in E(G)$, we have $\rho(\mathbf{A_{sys}}(G_{uv}))< \rho(\mathbf{A_{sys}}(G))$.
\end{corollary}

\begin{IEEEproof}
    Since $G$ is not regular and $\delta(G)\geq 2$, it follows from Theorem~\ref{thm:pfthm} (iv) that $\rho(\mathbf{A_{sys}}(G))> 1$. Similar to the proof of Theorem~\ref{thm:add edge}, we can prove that there exists a vector $\boldsymbol{x}>0$ such that $\mathbf{A_{sys}}(G_{uv})\boldsymbol{x}< \rho(\mathbf{A_{sys}}(G))\boldsymbol{x}$. Therefore, by Theorem~\ref{thm:pfthm} (iii), we conclude that $\rho(\mathbf{A_{sys}}(G_{uv}))< \rho(\mathbf{A_{sys}}(G))$.
\end{IEEEproof}

\begin{remark}\label{remark:radius}
    The theta graph $\theta(l_1,l_2,l_3)$ satisfies the conditions in Corollary~\ref{cor:radius}. Therefore, subdividing an edge $uv$ reduces the spectral radius of the system matrix, implying that $\rho (A(\theta(l_1,l_2,l_3)))>\rho (A(\theta(l_1+1,l_2,l_3)))$ (similarly for $l_2$ and $l_3$).
    Similarly, for the dumbbell graph $D_b(r_1,r_2;q)$, we have $\rho (A(D_b(r_1,r_2;q)))>\rho (A(D_b(r_1+1,r_2;q)))$ (similarly for $r_2$) and $\rho (A(D_b(r_1,r_2;q)))>\rho (A(D_b(r_1,r_2;q+1)))$ for $q\geq 1$.
\end{remark}

We also compute specific examples for different cycle lengths and distances, with results summarized in Table~\ref{tab:radius}. The first column lists different distances between cycles, while the first row shows various combinations of cycle lengths. The symbol `-' indicates that no graph structure corresponds to the given distance and combination of cycles. 
For example, `$C_3,C_3$' with a distance of `$-1$' corresponds to $\theta(1,2,2)$, and a distance of `$1$' corresponds to $D_b(3,3;1)$. 
Since the distance between two cycles of lengths $c_1$ and $c_2$ satisfies $d\geq \max\{-\left\lfloor \frac{c_1}{2}\right\rfloor, -\left\lfloor \frac{c_2}{2}\right\rfloor\}$, there is no graph corresponding to `$C_3 , C_3$' with a distance of `$-2$'. 

\begin{table}[!tb]
    \setlength\tabcolsep{1pt}
    \begin{center}
    \caption{The spectral radius and the corresponding graphs for different cycle lengths and distances between cycles.}
    \label{tab:radius}
    \begin{tabular}{|c|c|c|c|c|c|c|}
    \hline
    $\rho$ & $C_3,C_3$ & $C_3 , C_4$ & $C_3 , C_5$ & $C_4 , C_4$ & $C_4 , C_5$ & $C_5 , C_5$\\ \hline
    -2 & - & - & - & \begin{tabular}[c]{@{}c@{}}1.4142\\$\theta(2,2,2)$\end{tabular} & \begin{tabular}[c]{@{}c@{}}1.3479\\$\theta(2,2,3)$\end{tabular} & \begin{tabular}[c]{@{}c@{}}1.2980\\$\theta(2,3,3)$\end{tabular}\\ \hline
    -1 & \begin{tabular}[c]{@{}c@{}}1.5214\\$\theta(1,2,2)$\end{tabular} & \begin{tabular}[c]{@{}c@{}}1.4241\\$\theta(1,2,3)$\end{tabular} & \begin{tabular}[c]{@{}c@{}}1.3608\\$\theta(1,2,4)$\end{tabular} & \begin{tabular}[c]{@{}c@{}}1.3532\\$\theta(1,3,3)$\end{tabular} & \begin{tabular}[c]{@{}c@{}}1.3054\\$\theta(1,3,4)$\end{tabular} & \begin{tabular}[c]{@{}c@{}}1.2672\\$\theta(1,4,4)$\end{tabular}\\ \hline
    0 & \begin{tabular}[c]{@{}c@{}}1.4422\\$D_b(3,3;0)$\end{tabular} & \begin{tabular}[c]{@{}c@{}}1.3712\\$D_b(3,4;0)$\end{tabular} & \begin{tabular}[c]{@{}c@{}}1.3225\\$D_b(3,5;0)$\end{tabular} & \begin{tabular}[c]{@{}c@{}}1.3161\\$D_b(4,4;0)$\end{tabular} & \begin{tabular}[c]{@{}c@{}}1.2776\\$D_b(4,5;0)$\end{tabular} & \begin{tabular}[c]{@{}c@{}}1.2457\\$D_b(5,5;0)$\end{tabular}\\ \hline
    1 & \begin{tabular}[c]{@{}c@{}}1.3532\\$D_b(3,3;1)$\end{tabular} & \begin{tabular}[c]{@{}c@{}}1.3061\\$D_b(3,4;1)$\end{tabular} & \begin{tabular}[c]{@{}c@{}}1.2722\\$D_b(3,5;1)$\end{tabular} & \begin{tabular}[c]{@{}c@{}}1.2672\\$D_b(4,4;1)$\end{tabular} & \begin{tabular}[c]{@{}c@{}}1.2390\\$D_b(4,5;1)$\end{tabular} & \begin{tabular}[c]{@{}c@{}}1.2149\\$D_b(5,5;1)$\end{tabular}\\ \hline
    2 & \begin{tabular}[c]{@{}c@{}}1.2980\\$D_b(3,3;2)$\end{tabular} & \begin{tabular}[c]{@{}c@{}}1.2632\\$D_b(3,4;2)$\end{tabular} & \begin{tabular}[c]{@{}c@{}}1.2376\\$D_b(3,5;2)$\end{tabular} & \begin{tabular}[c]{@{}c@{}}1.2334\\$D_b(4,4;2)$\end{tabular} & \begin{tabular}[c]{@{}c@{}}1.2115\\$D_b(4,5;2)$\end{tabular} & \begin{tabular}[c]{@{}c@{}}1.1921\\$D_b(5,5;2)$\end{tabular}\\ \hline
    3 & \begin{tabular}[c]{@{}c@{}}1.2599\\$D_b(3,3;3)$\end{tabular} & \begin{tabular}[c]{@{}c@{}}1.2325\\$D_b(3,4;3)$\end{tabular} & \begin{tabular}[c]{@{}c@{}}1.2121\\$D_b(3,5;3)$\end{tabular} & \begin{tabular}[c]{@{}c@{}}1.2085\\$D_b(4,4;3)$\end{tabular} & \begin{tabular}[c]{@{}c@{}}1.1906\\$D_b(4,5;3)$\end{tabular} & \begin{tabular}[c]{@{}c@{}}1.1745\\$D_b(5,5;3)$\end{tabular}\\ \hline
    4 & \begin{tabular}[c]{@{}c@{}}1.2318\\$D_b(3,3;4)$\end{tabular} & \begin{tabular}[c]{@{}c@{}}1.2093\\$D_b(3,4;4)$\end{tabular} & \begin{tabular}[c]{@{}c@{}}1.1932\\$D_b(3,5;4)$\end{tabular} & \begin{tabular}[c]{@{}c@{}}1.1892\\$D_b(4,4;4)$\end{tabular} & \begin{tabular}[c]{@{}c@{}}1.1741\\$D_b(4,5;4)$\end{tabular} & \begin{tabular}[c]{@{}c@{}}1.1603\\$D_b(5,5;4)$\end{tabular}\\ \hline
    \end{tabular}
    \end{center}
\end{table}

\begin{figure}[!tb]
    \centering
    \includegraphics[width=2.5in]{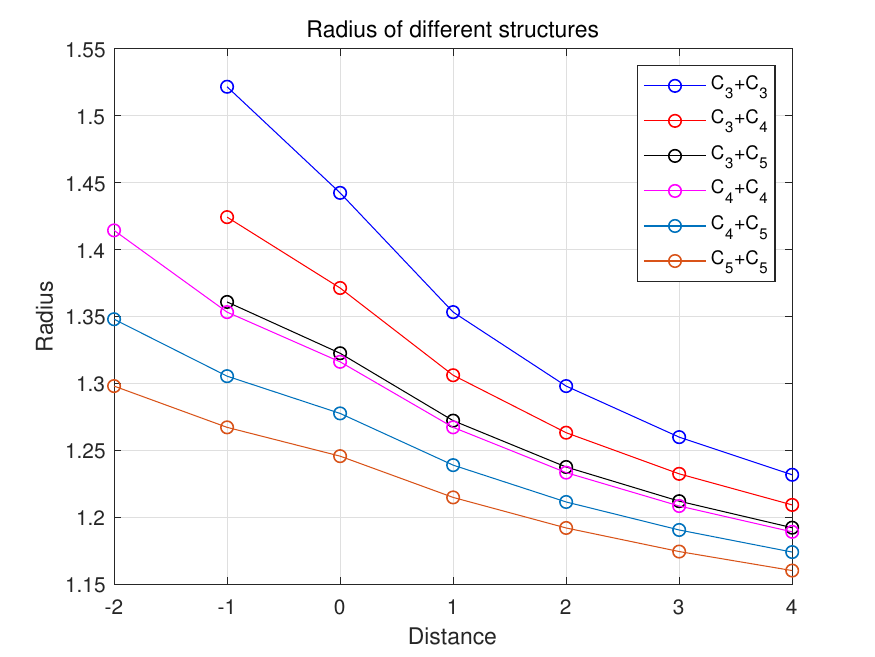}
    \caption{The variation of the spectral radius with respect to different cycle lengths and distances between cycles.}
    \label{fig:radius}
\end{figure}

The curve showing the variation of the spectral radius with different cycle lengths and distances is presented in Fig.~\ref{fig:radius}, which provides a more intuitive visualization. 

\begin{observation}\label{obs:radius variation}
    Based on Remark~\ref{remark:radius} and Fig.~\ref{fig:radius}, we notice that longer cycles and larger distances between cycles lead to smaller spectral radii. This implies that such structures are less harmful to the performance of LDPC codes. 
\end{observation}

\section{The Impact of Distances between Cycles on ETSs}\label{sec:6 cycles}

In this section, we focus on the impact of the distance between two 6-cycles in the Tanner graph on ETSs, which corresponds to the relationship between two 3-cycles in the VN graph. We consider the cases where the distances between the two 3-cycles are -1, 0, and 1, corresponding to theta graph $\theta(1,2,2)$, dumbbell graph $D_b(3,3;0)$ and $D_b(3,3;1)$, respectively. We compare the harmfulness of these three structures from two perspectives. First, we use the Tur\'an numbers of these graphs to calculate the ETSs eliminated when removing these structures. Second, we calculate the spectral radius of $(a, b)$-ETSs without these structures for various values of $a$ and $b$. Furthermore, we use the linear state-space model to analyze the rate of error probability reduction of these ETSs as the number of iterations increases. Specifically, we use density evolution to approximate the information outside the ETSs and the linear state-space model to study the information updating process in the ETSs.

\subsection{Tur\'an Numbers of These Graphs and ETSs}

By removing the theta graph $\theta(1,2,2)$ and the dumbbell graphs $D_b(3,3;0)$ and $D_b(3,3;1)$ in the VN graph of an $(a,b)$-ETS with variable degree $d_L(v)=\gamma$, we obtain the following relationships among the parameters $a$, $b$, and $\gamma$, directly from Theorem~\ref{thm:t122},~\ref{thm:d330}, and~\ref{thm:d331}.

\begin{corollary}\label{ets}
    For an $(a,b)$-ETS in a Tanner graph with variable-regular degree $d_L(v)=\gamma$, the parameters $a$, $b$, and $\gamma$ satisfy the following inequalities:
    \begin{itemize}
        \item[(1)] If there is no $\theta(1,2,2)$ in the VN graph, then $b\geq a\gamma-\frac{1}{2}a^2$;
        \item[(2)] If there is no $D_b(3,3;0)$ in the VN graph, then $b\geq a\gamma-\frac{1}{2}a^2-2$;
        \item[(3)] If there is no $D_b(3,3;1)$ in the VN graph, then $b\geq a\gamma-\frac{1}{2}(a+1)^2+2$.
    \end{itemize}
\end{corollary}

Based on Corollary~\ref{ets}, and considering the parity conditions of $a$, $b$, and $\gamma$, we provide the smallest values of $a$ for various values of $b$ when the above-mentioned theta graph and dumbbell graphs are removed from the VN graph of an $(a,b)$-ETS with $d_L(v)=\gamma$. These results are shown in Tables~\ref{tab1}. 
Computing the smallest value of $a$ using the Tur\'an numbers is less complicated compared to the computer enumeration method used in~\cite{amirzade2022protograph}.

\begin{table}[!tb]
    \begin{center}
    \caption{The smallest value of $a$ for various values of $b$ when the corresponding theta graphs and dumbbell graphs are removed from the VN graph.}
    \label{tab1}
    \renewcommand{\arraystretch}{1.2} 
    \begin{tabular}{|c|c|c|c|c|}
    \hline
    $\gamma$ & $b$ & $\theta(1,2,2)$-free & $D_b(3,3;0)$-free & $D_b(3,3;1)$-free \\ \hline
    \multirow{4}{*}{3} & 0 & 6 & 4 & 4 \\ \cline{2-5}
                       & 1 & 7 & 5 & 5 \\ \cline{2-5}
                       & 2 & 6 & 4 & 4 \\ \cline{2-5}
                       & 3 & 5 & 3 & 3 \\ \hline
    \multirow{4}{*}{4} & 0 & 8 & 8 & 5 \\ \cline{2-5}
                       & 2 & 8 & 7 & 5 \\ \cline{2-5}
                       & 4 & 7 & 6 & 4 \\ \cline{2-5}
                       & 6 & 6 & 6 & 6 \\ \hline
    \multirow{8}{*}{5} & 0 & 10 & 10 & 10 \\ \cline{2-5}
                       & 1 & 11 & 11 & 9 \\ \cline{2-5}
                       & 2 & 10 & 10 & 8 \\ \cline{2-5}
                       & 3 & 11 & 9 & 9 \\ \cline{2-5}
                       & 4 & 10 & 10 & 8 \\ \cline{2-5}
                       & 5 & 9 & 9 & 7 \\ \cline{2-5}
                       & 6 & 10 & 8 & 8 \\ \cline{2-5}
                       & 7 & 9 & 9 & 7 \\ \hline
    \end{tabular}
    \end{center}
\end{table}

Based on the results presented, we notice that removing structures with smaller distance between cycles in a Tanner graph can eliminate more ETSs.
From the perspective of Tur\'an numbers, we provide a theoretical explanation for improving the performance of LDPC codes in the error floor region by ensuring a larger distance between any two cycles in the Tanner graph. This insight provides a feasible direction for the design of LDPC codes.

\subsection{Spectral Radius of ETSs}

Based on Observation~\ref{obs:radius variation}, structures with smaller distances between cycles are more harmful to the performance of LDPC codes, and we have shown that $\rho(\theta(1,2,2))> \rho(D_b(3,3;0))> \rho(D_b(3,3;1))$. In this subsection, we focus on the spectral radii of ETSs when these structures are forbidden.
To isolate the impact of removing different structures, we calculate the spectral radius of the system matrix for each ETS in the following three disjoint sets and compute their median and mean values:
\begin{itemize}
    \item $\mathcal{S}_{\theta(1,2,2)-free}(a,b)$ represents all $(a,b)$-ETSs that contain $D_b(3,3;0)$ and $D_b(3,3;1)$ but are free of $\theta(1,2,2)$;
    \item $\mathcal{S}_{D_b(3,3;0)-free}(a,b)$ represents all $(a,b)$-ETSs that contain $\theta(1,2,2)$ and $D_b(3,3;1)$ but are free of $D_b(3,3;0)$;
    \item $\mathcal{S}_{D_b(3,3;1)-free}(a,b)$ represents all $(a,b)$-ETSs that contain $\theta(1,2,2)$ and $D_b(3,3;0)$ but are free of $D_b(3,3;1)$.
\end{itemize}

We use the computational graph theory tool `nauty'~\cite{nauty} to search for non-isomorphic VN graphs that satisfy our requirements. For each $(a,b)$-ETS with $d_L(v)=\gamma$, we search for an undirected graph on $a$ vertices, with $\frac{1}{2}(a\gamma-b)$ edges and each vertex having a degree within the range $[\lceil \frac{\gamma}{2} \rceil,\gamma]$. We then count the number of the three graphs $\theta(1,2,2)$, $D_b(3,3;0)$, and $D_b(3,3;1)$ in each non-isomorphic VN graph and classify their corresponding ETSs into the three sets $\mathcal{S}_{\theta(1,2,2)-free}(a,b)$, $\mathcal{S}_{D_b(3,3;0)-free}(a,b)$, and $\mathcal{S}_{D_b(3,3;1)-free}(a,b)$ accordingly. 

We calculate the spectral radius of $(a,b)$-ETSs with $\gamma=3$ and 4 for all $a\leq10$ and $b\leq4$, as shown in Table~\ref{tab:radius 3 regular} and Table~\ref{tab:radius 4 regular}. The following cases are omitted:
\begin{itemize}
    \item For $\gamma=3$, the dumbbell graph $D_b(3,3;0)$ is naturally avoided in all VN graphs. Thus, results for $\mathcal{S}_{D_b(3,3;0)-free}(a,b)$ are omitted. In this case, $\mathcal{S}_{\theta(1,2,2)-free}(a,b)$ represents $(a,b)$-ETSs containing $D_b(3,3;1)$ but free of $\theta(1,2,2)$, while $\mathcal{S}_{D_b(3,3;1)-free}(a,b)$ represents $(a,b)$-ETSs containing $\theta(1,2,2)$ but free of $D_b(3,3;1)$.
    \item Cases where $a\gamma-b$ is odd are omitted, as no such $(a,b)$-ETSs exist.
    \item Cases with $b=0$ are omitted, as all graphs are regular and $\rho=\gamma-1$ for all $(a,0)$-ETSs. This omission highlights the impact of removing different structures on the spectral radius.
\end{itemize}
In Table~\ref{tab:radius 4 regular}, the `-' symbol indicates that the corresponding set is empty for certain values of $a$ and $b$.

\begin{observation}\label{obs:radius comparison}
    For all the cases listed in Table~\ref{tab:radius 3 regular} and~\ref{tab:radius 4 regular}, $\rho_{mean}(\mathcal{S}_{\theta(1,2,2)-free}(a,b)) < \rho_{mean}(\mathcal{S}_{D_b(3,3;0)-free}(a,b))< \rho_{mean}(\mathcal{S}_{D_b(3,3;1)-free}(a,b))$. The same inequality holds for the median of the spectral radius. As mentioned in Section~\ref{sec:definition}, a larger spectral radius corresponds to a slower rate of error probability reduction. Therefore, removing $\theta(1,2,2)$ improves the performance of LDPC codes more than removing $D_b(3,3;0)$, and removing $D_b(3,3;0)$ is more effective than removing $D_b(3,3;1)$. In other words, $\theta(1,2,2)$ is more harmful to the performance than $D_b(3,3;0)$, and $D_b(3,3;0)$ is more harmful than $D_b(3,3;1)$.
\end{observation}

\begin{table}[!tb]
    \begin{center}
    \caption{The number of $(a,b)$-ETSs in $\mathcal{S}_{\theta(1,2,2)-free}(a,b)$ and $\mathcal{S}_{D_b(3,3;1)-free}(a,b)$, and the corresponding median and mean values of the spectral radius for various values of $a$, $b$, and $\gamma = 3$.}
    \label{tab:radius 3 regular}
    \begin{tabular}{|c|c|c|c|}
    \hline
    Sets with $\gamma=3$ & Num. & $\rho_{median}$ & $\rho_{mean}$\\ \hline
    $\mathcal{S}_{\theta(1,2,2)-free}(6,2)$ & 1 & 1.6956 & 1.6956\\
    $\mathcal{S}_{D_b(3,3;1)-free}(6,2)$      & 1 & 1.7293 & 1.7293\\ \hline
    $\mathcal{S}_{\theta(1,2,2)-free}(7,3)$ & 1 & 1.5986 & 1.5986\\
    $\mathcal{S}_{D_b(3,3;1)-free}(7,3)$      & 1 & 1.6653 & 1.6653\\ \hline
    $\mathcal{S}_{\theta(1,2,2)-free}(8,2)$ & 3 & 1.7822 & 1.7921\\
    $\mathcal{S}_{D_b(3,3;1)-free}(8,2)$      & 3 & 1.8250 & 1.8332\\ \hline
    $\mathcal{S}_{\theta(1,2,2)-free}(8,4)$ & 2 & 1.5451 & 1.5451\\
    $\mathcal{S}_{D_b(3,3;1)-free}(8,4)$      & 3 & 1.5990 & 1.6009\\ \hline
    $\mathcal{S}_{\theta(1,2,2)-free}(9,1)$ & 3 & 1.9116 & 1.9132\\
    $\mathcal{S}_{D_b(3,3;1)-free}(9,1)$      & 3 & 1.9189 & 1.9209\\ \hline
    $\mathcal{S}_{\theta(1,2,2)-free}(9,3)$ & 7 & 1.7185 & 1.7226\\
    $\mathcal{S}_{D_b(3,3;1)-free}(9,3)$      & 11 & 1.7471 & 1.7561\\ \hline
    $\mathcal{S}_{\theta(1,2,2)-free}(10,2)$ & 15 & 1.8371 & 1.8469\\
    $\mathcal{S}_{D_b(3,3;1)-free}(10,2)$      & 19 & 1.8461 & 1.8539\\ \hline
    $\mathcal{S}_{\theta(1,2,2)-free}(10,4)$ & 18 & 1.6598 & 1.6880\\
    $\mathcal{S}_{D_b(3,3;1)-free}(10,4)$      & 26 & 1.6916 & 1.7037\\ \hline
    \end{tabular}
    \end{center}
\end{table}

\begin{table}[!tb]
    \begin{center}
    \caption{The number of $(a,b)$-ETSs in $\mathcal{S}_{\theta(1,2,2)-free}(a,b)$, $\mathcal{S}_{D_b(3,3;0)-free}(a,b)$, and $\mathcal{S}_{D_b(3,3;1)-free}(a,b)$, and the corresponding median and mean values of the spectral radius for various values of $a$, $b$, and $\gamma = 4$.}
    \label{tab:radius 4 regular}
    \begin{tabular}{|c|c|c|c|}
    \hline
    Sets with $\gamma=4$ & Num. & $\rho_{median}$ & $\rho_{mean}$\\ \hline
    $\mathcal{S}_{\theta(1,2,2)-free}(7,4)$ & - & - & -\\
    $\mathcal{S}_{D_b(3,3;0)-free}(7,4)$      & 2 & 2.4680 & 2.4680\\
    $\mathcal{S}_{D_b(3,3;1)-free}(7,4)$      & 2 & 2.4745 & 2.4745\\ \hline
    $\mathcal{S}_{\theta(1,2,2)-free}(8,2)$ & - & - & -\\
    $\mathcal{S}_{D_b(3,3;0)-free}(8,2)$      & 5 & 2.7777 & 2.7792\\
    $\mathcal{S}_{D_b(3,3;1)-free}(8,2)$      & 1 & 2.7841 & 2.7841\\ \hline
    $\mathcal{S}_{\theta(1,2,2)-free}(8,4)$ & 2 & 2.5348 & 2.5348\\
    $\mathcal{S}_{D_b(3,3;0)-free}(8,4)$      & 11 & 2.5377 & 2.5365\\
    $\mathcal{S}_{D_b(3,3;1)-free}(8,4)$      & 2 & 2.5488 & 2.5488\\ \hline
    $\mathcal{S}_{\theta(1,2,2)-free}(9,2)$ & 4 & 2.8064 & 2.8061\\
    $\mathcal{S}_{D_b(3,3;0)-free}(9,2)$      & 15 & 2.8074 & 2.8115\\
    $\mathcal{S}_{D_b(3,3;1)-free}(9,2)$      & 2 & 2.8144 & 2.8144\\ \hline
    $\mathcal{S}_{\theta(1,2,2)-free}(9,4)$ & 16 & 2.5918 & 2.5927\\
    $\mathcal{S}_{D_b(3,3;0)-free}(9,4)$      & 52 & 2.5979 & 2.5984\\
    $\mathcal{S}_{D_b(3,3;1)-free}(9,4)$      & 12 & 2.6180 & 2.6209\\ \hline
    $\mathcal{S}_{\theta(1,2,2)-free}(10,2)$ & 39 & 2.8280 & 2.8290\\
    $\mathcal{S}_{D_b(3,3;0)-free}(10,2)$      & 116 & 2.8291 & 2.8320\\
    $\mathcal{S}_{D_b(3,3;1)-free}(10,2)$      & 6 & 2.8345 & 2.8353\\ \hline
    $\mathcal{S}_{\theta(1,2,2)-free}(10,4)$ & 127 & 2.6393 & 2.6412\\
    $\mathcal{S}_{D_b(3,3;0)-free}(10,4)$      & 387 & 2.6444 & 2.6454\\
    $\mathcal{S}_{D_b(3,3;1)-free}(10,4)$      & 41 & 2.6582 & 2.6677\\ \hline
    \end{tabular}
    \end{center}
\end{table}

Furthermore, to more intuitively show the impact of removing different structures, we use the linear state-space model to calculate the average rate of error probability reduction in the ETSs of the three sets $\mathcal{S}_{\theta(1,2,2)-free}(a,b)$, $\mathcal{S}_{D_b(3,3;0)-free}(a,b)$, and $\mathcal{S}_{D_b(3,3;1)-free}(a,b)$ for the case of $(10,4)$-ETSs with $\gamma=4$ as the number of iterations increases. 
For the external information from degree-1 check nodes, we use a Gaussian approximation for estimation, and we make the following assumptions, as done in~\cite{Butler2014errorfloor} and~\cite{schlegel2010dynamics}:
\begin{itemize}
    \item[(1)] The external inputs to the degree-2 check nodes in the ETS have a negligible effect on the information update within the ETS. 
    This assumption holds because the Gaussian approximation for estimating external information grows rapidly, so the information transmitted by the degree-2 check nodes is largely dominated by the information within the ETS.
    \item[(2)] The information generated within the ETS does not influence the external information updates. This is due to the fact that the number of nodes in the ETS is relatively small compared to the entire Tanner graph.
    \item[(3)] We use a Gaussian density function to approximate the probability density function of the messages. Additionally, we assume that the messages satisfy the consistency condition, meaning that the variance of the message's probability density function equals twice its mean. Thus, we only need to transmit the mean of the message during the process.
    \item[(4)] We assume that the information transmitted from the channel and the degree-1 check nodes in each iteration is statistically independent. Moreover, we assume that $\boldsymbol{\lambda_{ex}}^{(l)}$ between iterations is also independent.
\end{itemize}

Based on these assumptions, we calculate the average error probability of the set $\mathcal{S} _{\theta(1,2,2)-free}(10,4)$, $\mathcal{S}_{D_b(3,3;0)-free}(10,4)$, and $\mathcal{S}_{D_b(3,3;1)-free}(10,4)$ for $(10, 4)$-ETSs with $\gamma = 4$ at iteration $l$:
\begin{equation}
    Pr_{e}^{(l)}\{\mathcal{S}\}=\frac{1}{|\mathcal{S}|}\sum_{s\in \mathcal{S}}Pr_{e}^{(l)}\{s\},
\end{equation}
where $s$ represents an ETS in $\mathcal{S}$. For the error probability of an ETS, we estimate it by the maximum error probability of the variable nodes in the ETS:
\begin{equation}\label{eq:error vertex}
    Pr_{e}^{(l)}\{s\}=\max_{v\in s}Pr_{e}^{(l)}\{v<0\}.
\end{equation}
Based on assumption (3), we can rewrite equation (\ref{eq:error vertex}) as 
\begin{equation}
    Pr_{e}^{(l)}\{s\}=\max_{v\in s}Q(\frac{\boldsymbol{\tilde{\lambda}}^{(l)}(v)}{\sqrt{2|\boldsymbol{\tilde{\lambda}}^{(l)}(v)|}}),
\end{equation}
where $\boldsymbol{\tilde{\lambda}}^{(l)}$ is obtained by the linear state-space model (\ref{eq:linear state-space model}),
with $Q(x)$ being the Gaussian Q-function.

\begin{figure}[!tb]
    \centering
    \includegraphics[width=2.5in]{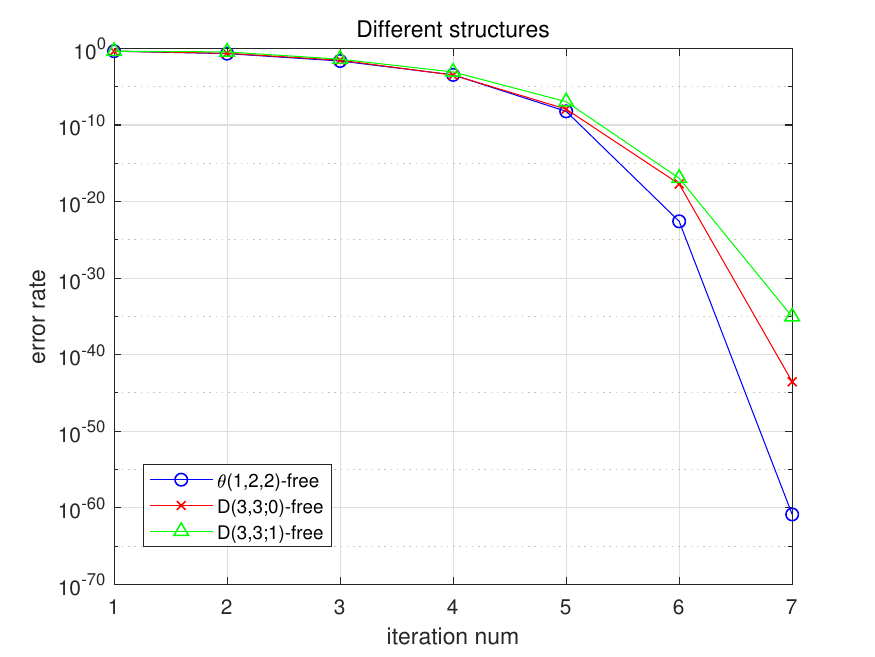}
    \caption{The average error probability of ETSs in $\mathcal{S}_{\theta(1,2,2)-free}(10,4)$, $\mathcal{S}_{D_b(3,3;0)-free}(10,4)$, and $\mathcal{S}_{D_b(3,3;1)-free}(10,4)$ as the number of iterations increases for $\gamma=4$.}
    \label{fig:error rate}
\end{figure}

We assume that the external Tanner graph 
corresponds to a (4,8)-regular LDPC code and that the channel variance is $\sigma=0.83$. The messages from the degree-1 check nodes are characterized by a Gaussian approximation. For the variable nodes in the ETS, we assume that the mean of the channel information is 0.01, indicating very poor information from the channel. This assumption allows us to analyze the rate at which the error probability decreases within the ETS.

When $\gamma=4$, as the number of iterations increases, the average error probabilities of the ETSs in $\mathcal{S}_{\theta(1,2,2)-free}(10,4)$, $\mathcal{S}_{D_b(3,3;0)-free}(10,4)$, and $\mathcal{S}_{D_b(3,3;1)-free}(10,4)$ are shown in Fig.~\ref{fig:error rate}, which further validate Observation~\ref{obs:radius comparison}.

\section{Constructions and Numerical Results}\label{sec:construction}

Based on the findings from previous sections, we modify the classical progressive-edge-growth (PEG) algorithm~\cite{hu2005regular}. At each step of adding an edge, we not only maximize the local cycle length but also ensure that the newly formed cycles are as far as possible from the existing cycles in the Tanner graph.

To achieve this, we assign a weight $wt(v_i)$ to each variable node $v_i$ to represent the cycles at $v_i$. Initially, the weights of all variable nodes are set to 0. Each time we perform a breadth-first expansion with $v_i$ as the root, we retain all paths from $v_i$ to each check node $c_j$, denoting these paths as $\mathcal{P}_{v_i}(c_j)=\{P\mid P\text{ is a path from }v_i\text{ to }c_j\}$. To save these paths, we check all connecting edges between adjacent layers in the expansion. Define the weight of $c_j$ in this expansion as the sum of the weights of all variable nodes in $\mathcal{P}_{v_i}(c_j)$:
\begin{equation}\label{eq:weight_of_chk}
    wt_{v_i}(c_j) = \sum_{P\in \mathcal{P}_{v_i}(c_j)} \sum_{\text{vaiable node }v\in P} wt(v)
\end{equation}
Note that some variable nodes may appear multiple times in different paths from $v_i$ to $c_j$ in the expansion. Therefore, when calculating the weight of $c_j$, these variable nodes will be counted multiple times.

When the expansion of $v_i$ that does not include all check nodes, we connect $v_i$ to the check node with the smallest degree that has not yet been included in the expansion, similar to the classical PEG algorithm. The difference arises when all check nodes have already been included in the expansion. In this case, we choose the check node $c_j$ at the deepest layer with the smallest weight to connect. During this process, new cycles are formed, as each path from $v_i$ to $c_j$ creates a new cycle. Consequently, we update the weights of all variable nodes in $\mathcal{P}_{v_i}(c_j)$. Specifically, we add a fixed value $w$ to the weight of each variable node. If a variable node appears $k$ times in $\mathcal{P}_{v_i}(c_j)$, it indicates that $k$ new cycles passing through this variable node are formed, so we add $k\times w$ to the weight of that variable node.

In this way, the weight of a variable node approximates the number of cycles passing through it, with a larger weight indicating more cycles. Therefore, we select the check node at the deepest layer with the smallest weight to connect. Choosing the deepest layer maximizes the local cycle length, while selecting the smallest weight ensures that the paths from the root to this check node intersect with the fewest cycles. This strategy guarantees that the newly formed cycle is as far as possible from existing cycles in the graph.

Note that the value of $w$ can be adjusted to further distinguish between cycles of different lengths. In this paper, we set $w$ as the reciprocal of the length of the newly formed cycle, to ensure that shorter cycles have larger weights than longer cycles. This prevents short cycles from being too dense in the Tanner graph. This process is further illustrated in the following example.

\begin{figure}[!tb]
    \centering
    \includegraphics[width=2.5in]{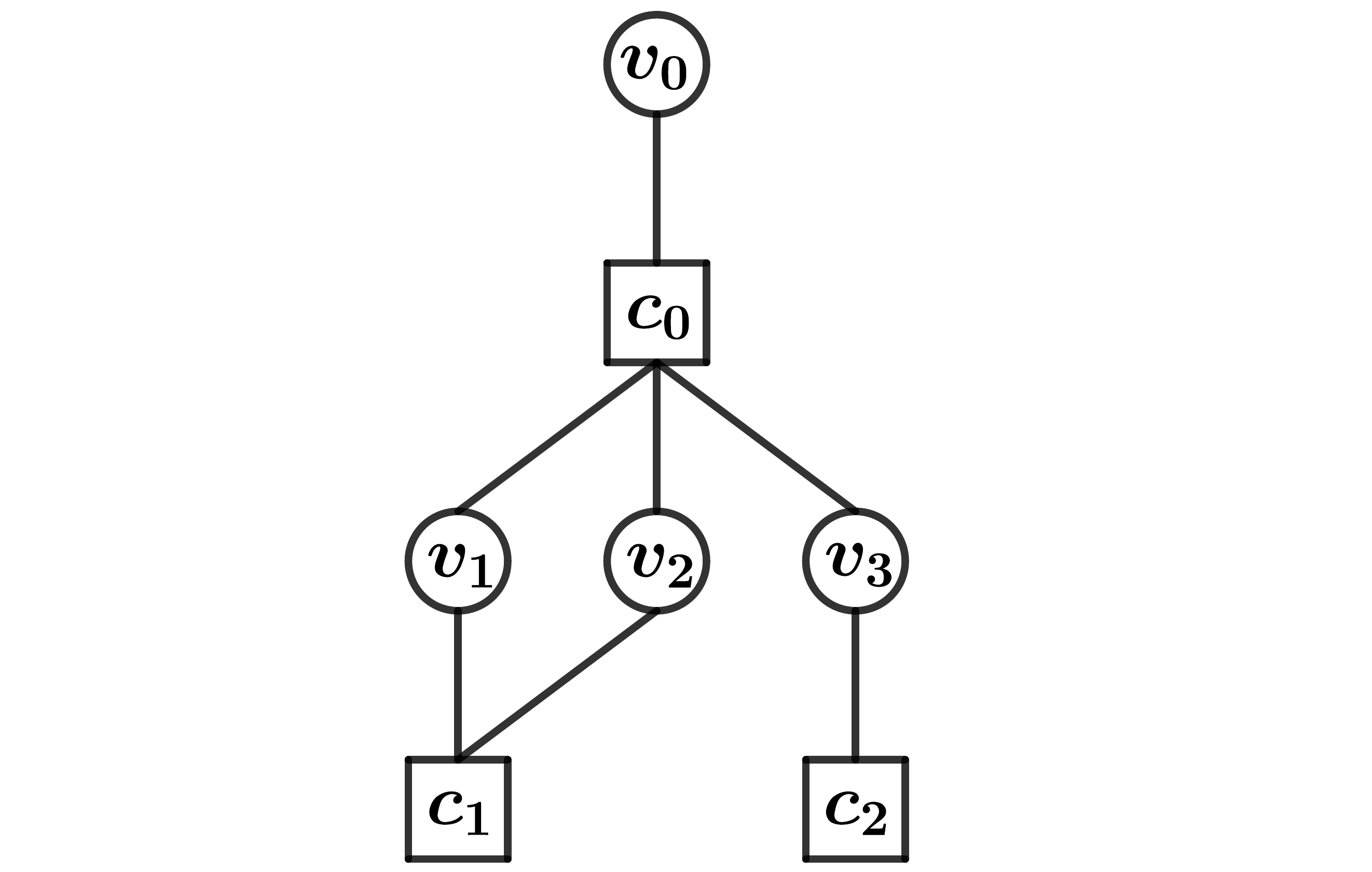}
    \caption{An expansion with variable node $v_0$ as the root, where circles represent variable nodes and squares represent check nodes.}
    \label{fig tree_expansion}
\end{figure}

\begin{example}
    For an expansion with $v_0$ as the root, as shown in Fig.~\ref{fig tree_expansion}, we have $\mathcal{P}_{v_0}(c_0) = \{(v_0c_0)\}$, $\mathcal{P}_{v_0}(c_1) = \{(v_0c_0v_1c_1),(v_0c_0v_2c_1)\}$, and $\mathcal{P}_{v_0}(c_2) = \{(v_0c_0v_3c_2)\}$. Then, $wt_{v_0}(c_1) = 2wt(v_0)+wt(v_1)+wt(v_2)$ and $wt_{v_0}(c_2) = wt(v_0)+wt(v_3)$. If $wt_{v_0}(c_1) < wt_{v_0}(c_2)$, we connect $v_0$ and $c_1$, forming 4-cycles, and set $w = \frac{1}{4}$. We then update the weights of the variable nodes as follows:
    \begin{itemize}
        \item The weight of $v_0$ is updated to $wt(v_0)+2\times \frac{1}{4}$;
        \item The weight of $v_1$ is updated to $wt(v_1)+ \frac{1}{4}$;
        \item The weight of $v_2$ is updated to $wt(v_2)+ \frac{1}{4}$.
    \end{itemize}
\end{example}

We design the PEG-CYCLE algorithm.

\begin{algorithm}[h]
	\caption{PEG-CYCLE algorithm}
	\begin{algorithmic}[1]
		\REQUIRE Number of variable nodes $n_v$, number of check nodes $n_c$, degree sequence of variable nodes $\{d_1, d_2, \dots, d_{n_v}\}$
		\ENSURE Parity-check matrix $\mathbf{H}$ of size $n_c \times n_v$
		\RETURN $\mathbf{H}$ as a zero matrix, $wt(v)=0$ for each variable node $v$
		\FOR{$i=1$ to $n_v$}
		    \FOR{$k=1$ to $d_i$}
                \IF{$k=1$}
                    \STATE Select the check node $c_j$ with the smallest degree (random selection if multiple options), update $\mathbf{H}(j, i) \gets 1$
                \ELSE
                    \STATE Perform breadth-first expansion with $v_i$ as the root, save paths $\mathcal{P}_{v_i}(c)$ for each check node $c$
                    \IF{there exist check nodes not included in the expansion}
                        \STATE Select the check node $c_j$ not yet in the expansion with the smallest degree (random selection if multiple options), update $\mathbf{H}(j, i) \gets 1$
                    \ELSE
                        \STATE Select the check node $c_j$ at the deepest layer with the smallest weight according to equation~(\ref{eq:weight_of_chk}) (random selection if multiple options), update $\mathbf{H}(j, i) \gets 1$
                        \FOR{Each $P\in\mathcal{P}_{v_i}(c_j)$}
                            \FOR{Each $v\in P$}
                                \STATE Update $wt(v) \gets wt(v) + w$
                            \ENDFOR
                        \ENDFOR
                    \ENDIF
                \ENDIF
            \ENDFOR
        \ENDFOR
	\end{algorithmic}\label{algo:peg}
\end{algorithm}

For QC-LDPC codes, the expansion is performed with the first variable node of each QC block as the root, and the selection of connected check nodes follows the rules outlined in Algorithm~\ref{algo:peg}. Based on this edge, all edges within the QC block are subsequently updated according to the quasi-cyclic structure. It is important to note that, due to the non-fully connected positions in the base matrix and the quasi-cyclic structure, some check nodes cannot be connected to the variable node. We refer to those that are able to add edges as available check nodes. After each edge addition, the available check nodes are updated. Furthermore, when updating the weights of each variable node, the same update must be applied to all variable nodes within the same QC block. We propose the following QC-PEG-CYCLE algorithm.

\begin{algorithm}[h]
	\caption{QC-PEG-CYCLE algorithm}
	\begin{algorithmic}[2]
		\REQUIRE Base matrix $\mathbf{B}$ of size $n_c \times n_v$, lifting degree $p$
		\ENSURE Parity-check matrix $\mathbf{H}$ of size $pn_c \times pn_v$
		\RETURN $\mathbf{H}$ as a zero matrix, $wt(v)=0$ for each variable node $v$, $\{d_1, d_2, \dots, d_{n_v}\}$ is the degree sequence of $\mathbf{B}$
		\FOR{$i=0$ to $n_v-1$}
		    \FOR{$k=1$ to $d_i$}
                \IF{$k=1$}
                    \STATE Select an available check node $c_j$ (random selection if multiple options), update $\mathbf{H}(j, 1+ip) \gets 1$
                \ELSE
                    \STATE Perform breadth-first expansion with $v_{1+ip}$ as the root, save paths $\mathcal{P}_{v_{1+ip}}(c)$ for each check node $c$
                    \IF{there exist available check nodes not included in the expansion}
                        \STATE Select an available check node $c_j$ not yet in the expansion (random selection if multiple options), update $\mathbf{H}(j, 1+ip) \gets 1$
                    \ELSE
                        \STATE Select the available check node $c_j$ at the deepest layer with the smallest weight according to equation~(\ref{eq:weight_of_chk}) (random selection if multiple options), update $\mathbf{H}(j, 1+ip) \gets 1$
                        \FOR{Each $P\in\mathcal{P}_{v_i}(c_j)$}
                            \FOR{Each $v\in P$}
                                \STATE Update $wt(v) \gets wt(v) + w$, update $wt(v') \gets wt(v') + w$ for all variable nodes $v'$ in the same QC block as $v$
                            \ENDFOR
                        \ENDFOR
                    \ENDIF
                \ENDIF
                \STATE Update all edges in the same QC block with $(j, 1+ip)$, update the available check nodes
            \ENDFOR
        \ENDFOR
	\end{algorithmic}\label{algo:qc peg}
\end{algorithm}

Next, we present the performance curves of the QC-LDPC codes constructed using Algorithm~\ref{algo:qc peg}. We compare our construction with the state-of-the-art constructions from~\cite{amirzade2023construction,karimi2019construction,karimi2020construction,5gnr}, as well as the classical PEG algorithm~\cite{hu2005regular}, which includes both fully connected and non-fully connected constructions. 
In this section, all codes are decoded using the sum-product algorithm (SPA) over an additive white Gaussian noise (AWGN) channel, paired with binary phase shift keying (BPSK) modulation. The maximum number of iterations is set to 20. The parameters of the simulated codes are given in Table~\ref{tab:PEG-CYCLE}.

\begin{table*}[!tb]
    \begin{center}
    \caption{The parameters of simulated codes.}
    \label{tab:PEG-CYCLE}
        \begin{tabular}{|c|c|c|c|c|}
            \hline
            Code & Type & Lifting degree $p$ & Length &Exponent matrix\\ \hline
            $C_1$ & (3,5)-regular & 27 & 135 & $\begin{bmatrix}
                        19 & 4 & 9 & 18 & 17\\
                        26 & 5 & 18 & 18 & 5\\
                        15 & 4 & 12 & 1 & 3\\
                    \end{bmatrix}$ \\\hline
            Chordless~\cite{amirzade2023construction} & (3,5)-regular & 27 & 135 & $\begin{bmatrix}
                0 & 0 & 0 & 0 & 0\\
                0 & 1 & 2 & 9 & 12\\
                0 & 3 & 8 & 23 & 21\\
            \end{bmatrix}$ \\\hline
            $C_2$ & (3,6)-regular & 80 & 480 & $\begin{bmatrix}
                        63 & 15 & 50 & 73 & 47 & 69\\
                        67 & 24 & 63 & 4 & 2 & 10\\
                        39 & 68 & 71 & 67 & 26 & 11\\
                    \end{bmatrix}$ \\\hline
            TS-free~\cite{karimi2019construction} & (3,6)-regular & 80 & 480 & $\begin{bmatrix}
                        0 & 0 & 0 & 0 & 0 & 0\\
                        0 & 7&39&41&45&61\\
                        0&35&43&51&66&36\\
                    \end{bmatrix}$ \\\hline
            $C_3$ & (4,8)-irregular & 72 & 576 & $\begin{bmatrix}
                45&\infty&\infty&5&21&51&71&37\\
                42&\infty&62&\infty&67&33&15&37\\
                \infty&24&13&\infty&35&70&7&9\\
                \infty&68&\infty&1&24&26&33&11\\
            \end{bmatrix}$\\\hline
            Irregular TS-free~\cite{karimi2020construction} & (4,8)-irregular & 72 & 576 & $\begin{bmatrix}
                0&\infty&\infty&0&0&0&0&0\\
                0&\infty&4&\infty&13&30&33&52\\
                \infty&17&5&\infty&39&12&52&47\\
                \infty&58&\infty&64&6&1&24&29\\
            \end{bmatrix}$ \\\hline
            $C_4$ & (4,10)-irregular & 64 & 640 & $\begin{bmatrix}
                36&0&5&50&\infty&\infty&40&\infty&\infty&38\\
                0&\infty&\infty&54&22&3&62&62&46&63\\
                0&1&\infty&38&26&\infty&\infty&\infty&5&\infty\\
                \infty&0&37&\infty&40&0&4&30&1&55\\
            \end{bmatrix}$\\\hline
            5G BG2~\cite{5gnr} & (4,10)-irregular & 64 & 640 & $\begin{bmatrix}
                9&53&12&26&\infty&\infty&61&\infty&\infty&13\\
                39&\infty&\infty&38&61&61&34&28&32&60\\
                17&50&\infty&44&52&\infty&\infty&\infty&48&\infty\\
                \infty&8&58&\infty&60&40&17&54&18&0\\
            \end{bmatrix}$ \\\hline
            \end{tabular}
    \end{center}
\end{table*}

\begin{figure}[!tb]
    \centering
    \includegraphics[width=2.5in]{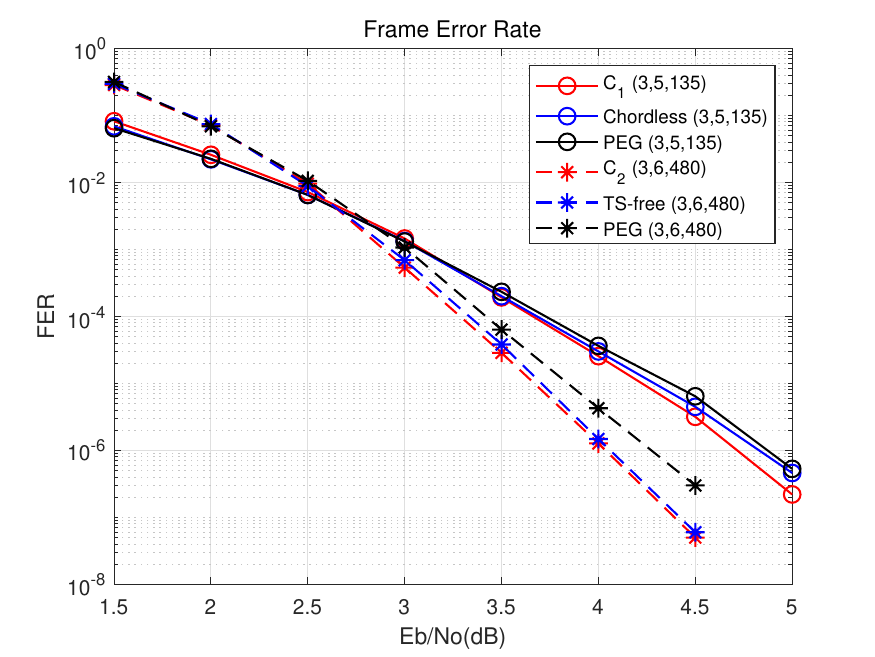}
    \caption{FER performance of $C_1$, $C_2$, and their counterparts for the fully connected case.}
    \label{fig:peg fully connected}
\end{figure}

Fig.~\ref{fig:peg fully connected} shows the performance of Algorithm~\ref{algo:qc peg} when applied to fully connected QC-LDPC codes. The codes $C_1$ and $C_2$ are generated using Algorithm~\ref{algo:qc peg}, and their performance is compared with the best-performing constructions from~\cite{amirzade2023construction} and~\cite{karimi2019construction}, denoted by `Chordless' and `TS-free', respectively. 
The construction `Chordless' is free of all cycles with a chord of length up to 12, while the construction `TS-free' is free of all $(a,b)$-ETSs with $a\leq 12$ and $b\leq 3$. Our constructions perform similarly to, or even better than, the best codes from~\cite{amirzade2023construction} and~\cite{karimi2019construction}. Furthermore, all four codes outperform those generated by the PEG algorithm.

\begin{figure}[!tb]
    \centering
    \includegraphics[width=2.5in]{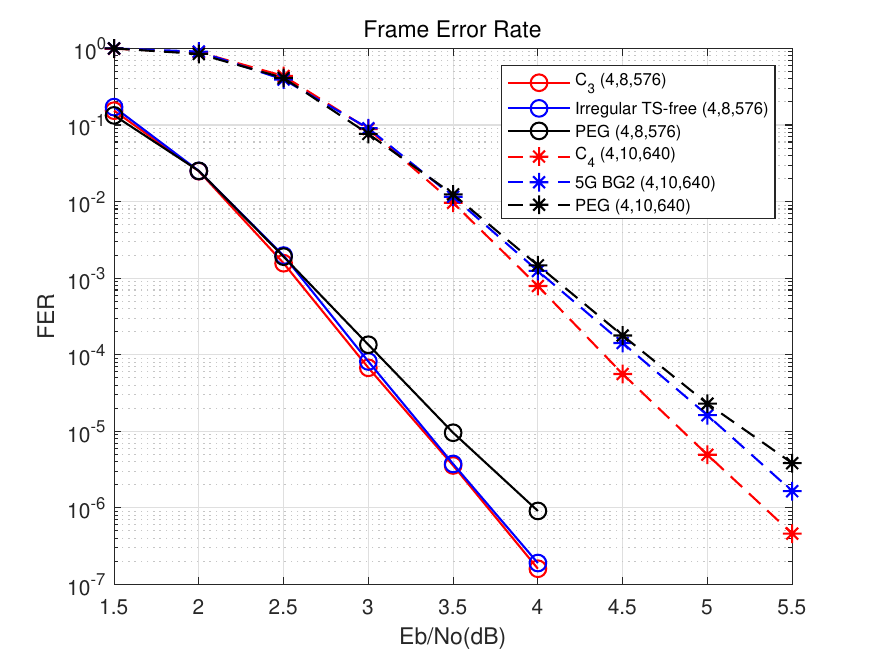}
    \caption{FER performance of $C_3$, $C_4$, and their counterparts for the non-fully connected case.}
    \label{fig:peg not fully connected}
\end{figure}

The performance of non-fully connected codes is shown in Fig.~\ref{fig:peg not fully connected}. The codes $C_3$ and $C_4$ are generated using Algorithm~\ref{algo:qc peg}, and their performance is compared with the best-performing constructions from~\cite{karimi2020construction} and the base matrix of 5G NR~\cite{5gnr}, denoted by `Irregular TS-free' and `5G BG2', respectively.  The construction `Irregular TS-free' focuses on removing all $(a,b)$-ETSs with $a\leq 8$ and $b\leq 3$, while `5G BG2' corresponds to the case of $p=64$. As shown in the figure, our constructions perform comparable to, or even superior to these state-of-the-art constructions. Moreover, all four codes outperform those generated by the PEG algorithm.

\section{Conclusion}\label{sec:conclusion}

In this paper, we study, for the first time, the impact of the distance between cycles on ETSs using theta graphs and dumbbell graphs. By determining two important graph parameters—the Tur\'an numbers and spectral radii—for these graphs, we observe that removing structures with smaller distance between cycles can eliminate more ETSs in the Tanner graph. Moreover, a larger distance between cycles corresponds to a smaller spectral radius of the system matrix in linear state-space model, thereby improving the performance of LDPC codes. 

We focus on three specific cases: two 6-cycles with distances of -1, 0, and 1 in the Tanner graph, represented by $\theta(1,2,2)$, $D_b(3,3;0)$, and $D_b(3,3;1)$ in the VN graph, respectively. Using their Tur\'an numbers, we compute the ETSs eliminated by removing these graphs. Additionally, we analyze all $(10,4)$-ETSs with $d_L(v)=4$, calculating the average spectral radius and error probability for sets free of $\theta(1,2,2)$, $D_b(3,3;0)$, and $D_b(3,3;1)$. These results are consistent with our findings. We also design the PEG-CYCLE algorithm, which greedily maximizes the distance between cycles in the Tanner graph. The proposed algorithm shows improved performance for both fully connected and non-fully connected QC-LDPC codes, outperforming the classical PEG algorithm and achieving performance comparable to or better than state-of-the-art construction methods.

\begin{appendices}

    \section{Proof of Theorem~\ref{thm:odd_even}}\label{appendix A}
    
    We first prove some necessary lemmas.

    \begin{lemma}\label{lemma:minimum-degree}
        Let $t$, $k$, and $n$ be integers with $t\geq 2$, $k\geq0$, and $n\geq \left\lfloor \frac{(t+1)^2}{4}\right\rfloor$. If $G$ is a graph on $n$ vertices and $|E(G)|\geq \left\lfloor \frac{n^2}{4}\right\rfloor +k$, then there exists an induced subgraph $G'\subseteq G$ on $n'\geq t$ vertices such that $|E(G')|\geq \left\lfloor \frac{n'^2}{4}\right\rfloor +k$ edges and $\delta (G')\geq \left\lfloor \frac{n'}{2}\right\rfloor$.
    \end{lemma}
    
    \begin{IEEEproof}
        The proof follows a similar strategy to~\cite[Lemma 4]{zhai2021turan}. For convenience, we outline the key steps here. The main idea of proof is iteratively deleting the vertex with the minimum degree.
        
        Initially, let $G=G_0$. If $\delta(G_0)\geq \lfloor \frac{n}{2}\rfloor$, the proof is complete. Otherwise, there exists a vertex $v_0$ in $G_0$ with degree $d_{G_0}(v_0)\leq \lfloor \frac{n}{2}\rfloor-1$. Let $G_1=G_0-v_0$, so $G_1$ contains $n-1$ vertices and satisfies $|E(G_1)|\geq \left\lfloor \frac{(n-1)^2}{4}\right\rfloor +k+1$. If $\delta(G_1)\geq \lfloor \frac{n-1}{2}\rfloor$, the process terminates. Otherwise, we can find a vertex $v_1$ with $d_{G_1}(v_1)\leq \lfloor \frac{n-1}{2}\rfloor-1$. Deleting $v_1$ from $G_1$ yields $G_2=G_1-v_1$, which has $n-2$ vertices and at least $\left\lfloor \frac{(n-2)^2}{4}\right\rfloor +k+2$ edges. This procedure repeats until we obtain the desired graph $G_i=G'$ for some $i\leq n-t$. 
        Otherwise, through this process, we construct a sequence of graphs $G_0,G_1,\dots,G_{n-t}$ such that $|E(G_i)|\geq \left\lfloor \frac{(n-i)^2}{4}\right\rfloor +k+i$ and $\delta(G_i)\leq \lfloor \frac{n-i}{2}\rfloor-1$. Consider $G_{n-t}$, which has $t$ vertices and at least $\left\lfloor \frac{t^2}{4}\right\rfloor +k+n-t$ edges. This is greater or equal to $\frac{t(t-1)}{2}=|E(K_t)|$ when $n\geq \lfloor \frac{(t+1)^2}{4}\rfloor$, creating a contradiction with $\delta(G_{n-t}) \leq \lfloor \frac{t}{2}\rfloor-1$. 
    \end{IEEEproof}
    
    Next lemma allows us to extend the path of length 1 in $D_b(3,k;1)$ to a path of length $q$:
    
    \begin{lemma}\label{lemma:d3kc}
        Given two fixed integers $q\geq2$ and $k\geq 3$, let $G$ be a graph on $n\geq 4k+4q-4$ vertices with $\delta(G)\geq \lfloor \frac{n}{2}\rfloor$. If $G$ contains a $D_b(3,k;1)$, then $G$ also contains a copy of $D_b(3,k;i)$ for all $2\leq i\leq q$.
    \end{lemma}
    
\begin{figure}[!tb]
        \centering
        \includegraphics[width=2.5in]{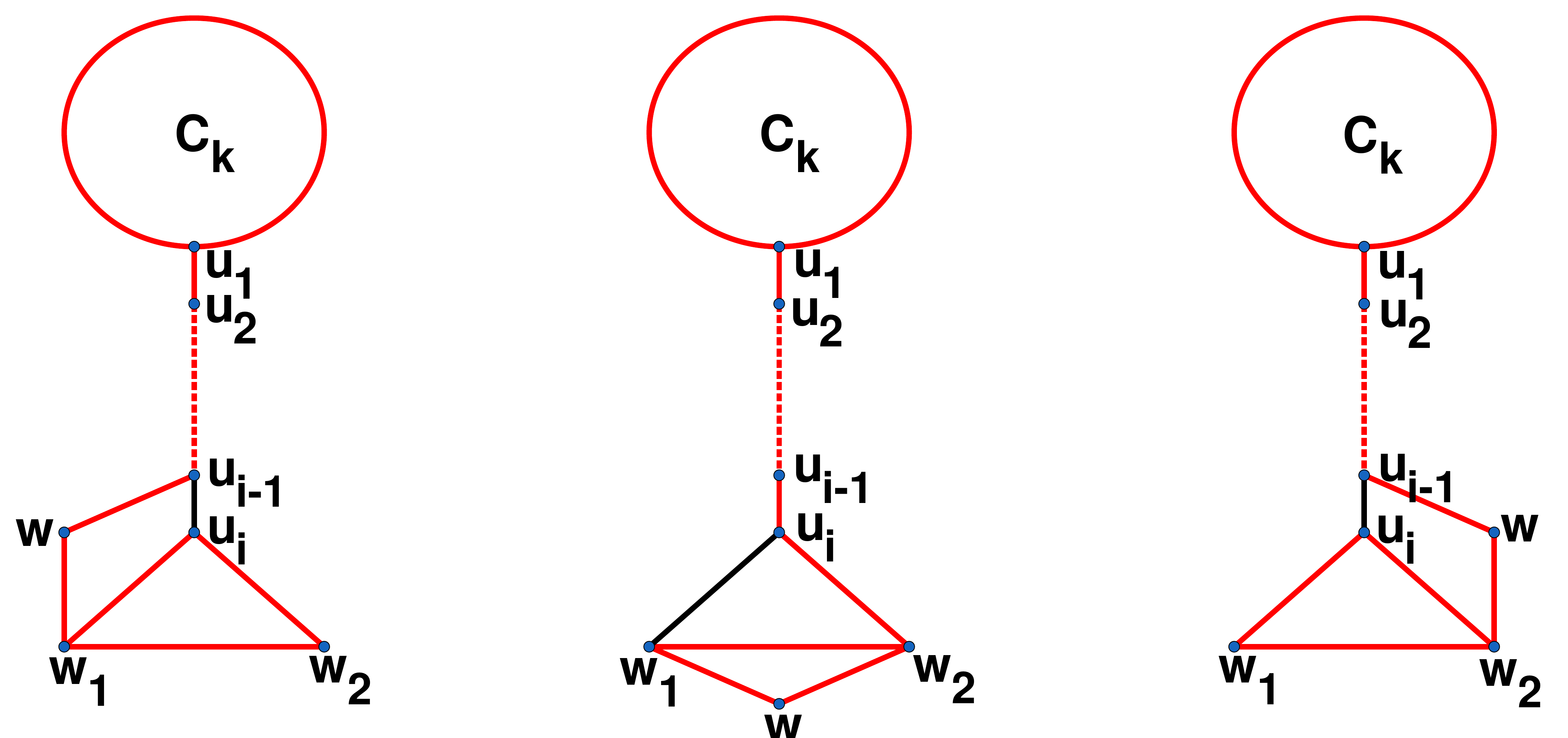}
        \caption{A $D_b(3,k;i)$ in $G$ labeled with the red color.}
        \label{fig1}
    \end{figure}

    \begin{IEEEproof}
        We prove this lemma by induction. Suppose $G$ contains a $D_b(3,k;i-1)$. We will show that $G$ also contains a $D_b(3,k;i)$ for all $2\leq i\leq q$.
    
        Denote the vertices of $D_b(3,k;i-1)$ as $V(D_b(3,k;i-1))=\{v_1,v_2,\dots,v_{k-1},u_1,u_2,\dots,u_{i},w_1,w_2\}$,
        where $\{v_1,v_2,\dots,v_{k-1},u_1\}$ forms the cycle $C_k$, $\{u_i,w_1,w_2\}$ forms the cycle $C_3$, and $\{u_1u_2\dots u_{i}\}$ is a path of length $i-1$ connecting $C_k$ and $C_3$.
        As
        \begin{eqnarray*}
            &&\sum_{x\in \{w_1,w_2,u_{i-1}\}}|N_G(x)-V(D_b(3,k;i-1))|\\
            &\geq &3(\delta(G)-(k+i))+3\\
            &\geq& 3\lfloor \frac{n}{2}\rfloor-3(k+i)+3 \\
            &> &n-(k+i+1)\\
            &=&|V(G)- V(D_b(3,k;i-1))|
        \end{eqnarray*}
        by $n\geq 4k+4q-4\geq 4k+4i-4$, there exists at least one vertex $w$ in $V(G)- V(D_b(3,k;i-1))$ that is adjacent to at least two vertices from $\{w_1,w_2,u_{i-1}\}$. 
        Once such a vertex $w$ is found, we can add it to the existing structure of $D_b(3,k;i-1)$, thereby obtaining a $D_b(3,k;i)$ (see Fig.~\ref{fig1}).
    \end{IEEEproof}
    
    Having extended the path in the dumbbell graph, we now use the following lemma from~\cite{zhai2021turan} to expand the cycles:
    
    \begin{lemma}[\cite{zhai2021turan}]\label{lemma:path-extension}
        Let $l$, $m$, $n$ be three positive integers with $l\geq 3$, $m\geq 2$, and $n\geq 4l+12m-18$. Let $G$ be a graph on $n$ vertices with $\delta(G)\geq \lfloor \frac{n}{2}\rfloor$, and $H$ be a subgraph of $G$ with $|V(H)|=l$. If $u_0w_1v_0$ is a path in $H$, then $G$ contains a $(u_0,v_0)$-path $P$ of length $2m$ such that $V(P)\cap V(H)\subseteq \{u_0,w_1,v_0\}$.
    \end{lemma}
    
    We now obtain the following lemma, which extends the graph $D_b(x,y;q)$ to $D_b(r_1,r_2;q)$ where $r_1\geq x$ and $r_2\geq y$.
    
    \begin{lemma}\label{lemma:cycle-extension}
        Let $x$, $y$, $r_1$, $r_2$, $q$ be five positive integers with $r_1\geq x\geq 3$, $r_2\geq y\geq 3$,  $q\geq 0$, and $r_1$, $x$ sharing the same parity, as well as $r_2$ and $y$. Let $G$ be a graph on $n$ vertices with $\delta(G)\geq \lfloor \frac{n}{2}\rfloor$. If $n\geq \min\{k_1,k_2\}$ and $G$ contains a $D_b(x,y;q)$, where
        \begin{equation*}
            k_1=\max\{-2x+4y+4q+6r_1-10,-2y+4q+4r_1+6r_2-10\},
        \end{equation*}
        \begin{equation*}
            k_2=\max\{-2y+4x+4q+6r_2-10,-2x+4q+4r_2+6r_1-10\},
        \end{equation*}
        then $G$ also contains a $D_b(r_1,r_2;q)$.
    \end{lemma}
    
    \begin{IEEEproof}
        We begin by applying Lemma~\ref{lemma:path-extension} twice: the first application to the $x$-cycle and the second to the $y$-cycle in $D_b(x,y;q)$. This extension process yields an $r_1$-cycle and an $r_2$-cycle, respectively, resulting in a $D_b(r_1,r_2;q)$.
    
        More specifically, denote the vertex set by
            $V(D_b(x,y;q))=\{u_1,u_2,\dots,u_x,w_1,w_2,\dots,w_{q-1},v_1,v_2,\dots,v_y\}$,
        where $\{u_1,u_2,\dots,u_x\}$ forms the $x$-cycle, $\{v_1,v_2,\dots,v_y\}$ forms the $y$-cycle, and $\{u_1,w_1,w_2,\dots,w_{q-1},v_1\}$ is the path of length $q$ connecting these two cycles.
        
        In the case where $n\geq k_1$, let us denote $H_1=D_b(x,y;q)$ and set $|V(H_1)|=l_1=x+y+q-1$, and $2m_1=r_1-x+2$. Since
            $n\geq 4l_1+12m_1-18=-2x+4y+4q+6r_1-10$
        and because $u_1u_2u_3$ is a path in $H_1$, Lemma~\ref{lemma:path-extension} guarantees that $G$ contains a $(u_1,u_3)$-path $P_1$ of length $2m_1=r_1-x+2$ such that $V(P_1)\cap V(H_1)\subseteq \{u_1,u_2,u_3\}$. Therefore, $H_1\cup P_1$ forms a $D_b(r_1,y;q)$, which we denote as $H_2$.
        
        Similarly, we can set $2m_2=r_2-y+2$, $|V(H_2)|=l_2=r_1+y+q-1$, and apply Lemma~\ref{lemma:path-extension} to the path $v_1v_2v_3$ in $H_2$. Since
            $n\geq 4l_2+12m_2-18=-2y+4q+4r_1+6r_2-10$,
        $G$ must contain a $(v_1,v_3)$-path $P_2$ of length $2m_2=r_2-y+2$ such that $V(P_2)\cap V(H_2)\subseteq \{v_1,v_2,v_3\}$. Then $H_2\cup P_2$ yields a $D_b(r_1,r_2;q)$.
        
        For the case where $n\geq k_2$, we first obtain a $D_b(x,r_2;q)$ from $D_b(x,y;q)$ and subsequently derive a $D_b(r_1,r_2;q)$. This completes the proof.
    \end{IEEEproof}

    We provide a brief overview of our proof framework: starting from the base cases of $D_b(3,3;0)$ and $D_b(3,3;1)$, we use Lemma~\ref{lemma:path-extension} to extend the path in the dumbbell graph and use Lemma~\ref{lemma:cycle-extension} to expand the cycles at both ends, thus completing the proof of the main theorem.

    Note that the cycle expansion process maintains the parity of their lengths. This implies the necessity of using 3-cycles and 4-cycles to generate larger odd and even cycles, respectively. To utilize Lemma~\ref{lemma:cycle-extension} for determining $ex(n,D_b(r_1,r_2;q))$ when $r_1+r_2$ is odd, we need the following basic results for $D_b(3,4;0)$ and $D_b(3,4;1)$:

\begin{lemma}\label{lemma:d34i}
    For an integer $i\in\{0,1\}$, let $G$ be a graph on $n\geq 16+4i$ vertices and $\delta(G)\geq \lfloor \frac{n}{2}\rfloor$. If $G$ contains a $C_3$, then $G$ also contains a copy of $D_b(3,4;i)$.
\end{lemma}

\begin{IEEEproof}
    Denote the vertices of the $C_3$ as $\{v_1,v_2,v_3\}$. 
    For $i=0$, consider the neighbors of $v_3$. We aim to show that a $C_4$ exists in $V(G)-\{v_1,v_2\}$ that contains $v_3$. Given that $d_G(v_3)\geq \delta(G)\geq \lfloor \frac{n}{2}\rfloor$, we can choose $\{v_4,v_5,v_6\}$ from $N_G(v_3)-\{v_1,v_2\}$.
    Note that: $\sum_{x\in \{v_4,v_5,v_6\}}|N_G(x)-\{v_1,v_2,\dots,v_6\}|\geq 3(\delta(G)-5)+3 \geq 3\lfloor \frac{n}{2}\rfloor-12$.
    Since $n\geq 16+4i$, it follows: $3\lfloor \frac{n}{2}\rfloor-12>n-6$.
    By the Pigeonhole principle, there must be a vertex $v_7$ in $V(G)-\{v_1,v_2,\dots,v_6\}$ that is adjacent to at least two vertices from $\{v_4,v_5,v_6\}$, forming a $D_b(3,4;0)$. 

    For $i=1$, apart from $v_1$ and $v_2$, we select another neighbor of $v_3$, denoted $v_0$, and then consider the neighbors of $v_0$. Similarly to the case of $i=0$, we can find a $C_4$ in $V(G)-\{v_1,v_2,v_3\}$ that contains $v_0$ when $n\geq 16+4i$, leading the presence of a $D_b(3,4;1)$. 
\end{IEEEproof}

Based on the foundational cases and the cycle expansion process, we can prove the following:

\begin{IEEEproof}[Proof of Theorem~\ref{thm:odd_even}]
    For the complete bipartite graph $K_{\lfloor \frac{n}{2}\rfloor,\lceil \frac{n}{2}\rceil}$, it is obvious that there are no odd cycles, so 
        $ex(n,D_b(r_1,r_2;q))\geq |E(K_{\lfloor \frac{n}{2}\rfloor,\lceil \frac{n}{2}\rceil})|=\lfloor \frac{n^2}{4}\rfloor$.
    Assume that $G$ is a graph on $n\geq k^2+k=\lfloor \frac{(2k+1)^2}{4}\rfloor$ vertices with $\lfloor \frac{n^2}{4}\rfloor+1$ edges. By Lemma~\ref{lemma:minimum-degree}, there exists an induced subgraph $G'$ on $n'\geq 2k=6r_1+6r_2+4q-24$ vertices with $\lfloor \frac{n'^2}{4}\rfloor+1$ edges and $\delta(G')\geq \lfloor \frac{n'}{2}\rfloor$. 
    
    By Theorem~\ref{thm:mantel}, there exists a $C_3$ in $G'$. Applying Lemma~\ref{lemma:d34i}, we find a $D_b(3,4;i)$ in $G'$ with $i\in\{0,1\}$. For the case where $q=0$, we use $D_b(3,4;0)$ to obtain $D_b(r_1,r_2;0)$ through Lemma~\ref{lemma:cycle-extension}. For the case where $q>0$, we apply Lemma~\ref{lemma:d3kc} and Lemma~\ref{lemma:cycle-extension} to extend the paths and cycles, respectively, finally obtaining $D_b(r_1,r_2;q)$ in $G'$.
    Therefore, we conclude that
        $ex(n,D_b(r_1,r_2;q))\leq \lfloor \frac{n^2}{4}\rfloor$.
\end{IEEEproof}

    \section{Proof of Theorem~\ref{thm:odd_odd_1}}\label{appendix B}

    \begin{itemize}
        \item[(i)] When $q=0$, to establish the lower bound, consider a graph $G_0$ on $n$ vertices whose edge set $E(G_0)=E(K_{\lfloor \frac{n}{2}\rfloor,\lceil \frac{n}{2}\rceil})\cup \{xy\}$, where $x$ and $y$ are non-adjacent vertices in $K_{\lfloor \frac{n}{2}\rfloor,\lceil \frac{n}{2}\rceil}$. Since each odd cycle in $G_0$ must contain the edge $e=xy$, it follows that no $D_b(r_1,r_2;0)$ exists in $G_0$. Thus, we have
            $ex(n,D_b(r_1,r_2;0))\geq \lfloor \frac{n^2}{4}\rfloor+1$.
        
        For the upper bound, 
        consider a graph $G$ on $n$ vertices with $|E(G)|\geq \lfloor \frac{n^2}{4}\rfloor+2$. By Lemma~\ref{lemma:minimum-degree}, there exists an induced subgraph $G'$ of $G$ with $n'\geq 2k=6r_1+6r_2-22$ vertices, $|E(G)|\geq \lfloor \frac{n^2}{4}\rfloor+2$ edges, and $\delta(G')\geq \lfloor \frac{n'}{2}\rfloor$. 
        By Theorem~\ref{thm:d330}, a $D_b(3,3;0)$ is found in $G'$. Then by Lemma~\ref{lemma:cycle-extension}, $G'$ contains a $D_b(r_1,r_2;0)$, as well as $G$.
        Therefore, we obtain
            $ex(n,D_b(r_1,r_2;0))\leq \lfloor \frac{n^2}{4}\rfloor+1$.
        \item[(ii)] When $q\geq1$, consider the graph $G_0$ on $n$ vertices with $E(G_0)=E(K_{\lfloor \frac{n}{2}\rfloor,\lceil \frac{n}{2}\rceil}) \cup E(K_{1,\lceil \frac{n}{2}\rceil-1})$, where $K_{1,\lceil \frac{n}{2}\rceil-1}$ is a star in the same part of $K_{\lfloor \frac{n}{2}\rfloor,\lceil \frac{n}{2}\rceil}$. Clearly, any two odd cycles in $G_0$ must share at least one common vertex, implying the absence of $D_b(r_1,r_2;q)$ in $G_0$. This leads to
            $ex(n,D_b(r_1,r_2;q))\geq |E(G_0)|=\lfloor \frac{n^2}{4}\rfloor+\lceil \frac{n}{2}\rceil-1$.

        For the upper bound, let $G$ be a graph on $n\geq k^2+k=\lfloor \frac{(2k+1)^2}{4}\rfloor$ vertices with $\lfloor \frac{n^2}{4}\rfloor+\lceil \frac{n}{2}\rceil$ edges. By Lemma~\ref{lemma:minimum-degree}, there exists an induced subgraph $G'$ on $n'\geq 2k=6r_1+6r_2+4q-22$ vertices, with $|E(G)|\geq \lfloor \frac{n'^2}{4}\rfloor+\lceil \frac{n}{2}\rceil \geq \lfloor \frac{n'^2}{4}\rfloor+\lceil \frac{n'}{2}\rceil$ edges, and $\delta(G')\geq \lfloor \frac{n'}{2}\rfloor$. 
        By Theorem~\ref{thm:d331}, there is a $D_b(3,3;1)$ in $G'$. By Lemma~\ref{lemma:d3kc}, we find a $D_b(3,3;q)$ in $G'$. Through Lemma~\ref{lemma:cycle-extension}, $G'$ contains a $D_b(r_1,r_2;q)$, as well as $G$.
        Thus, we have $ex(n,D_b(r_1,r_2;q))\leq \lfloor \frac{n^2}{4}\rfloor+\lceil \frac{n}{2}\rceil-1$.
    \end{itemize}

\end{appendices}

\bibliography{reference.bib}

\bibliographystyle{IEEEtran}

\end{document}